\documentclass[fleqn,usenatbib]{mnras}

\usepackage{newtxtext,newtxmath}

\usepackage[T1]{fontenc}

\DeclareRobustCommand{\VAN}[3]{#2}
\let\VANthebibliography\thebibliography
\def\thebibliography{\DeclareRobustCommand{\VAN}[3]{##3}\VANthebibliography}

\usepackage{graphicx}	
\usepackage{amsmath}	
\usepackage[dvipsnames]{xcolor}
\usepackage{comment}

\newcommand{\hMsun}{h^{-1}\mathrm{M_\odot}}
\newcommand{\hMpc}{h^{-1}\mathrm{Mpc}}
\newcommand{\invhMpc}{h\mathrm{Mpc}^{-1}}
\newcommand{\hGpc}{h^{-1}\mathrm{Gpc}}

\newcommand{\vect}[1]{\boldsymbol{#1}}
\newcommand{\fsig}{f\sigma_8}
\newcommand{\apar}{\alpha_\parallel}
\newcommand{\aperp}{\alpha_\bot}
\newcommand{\LCDM}{$\Lambda$CDM}
\newcommand{\dH}{D_\mathrm{H}}
\newcommand{\dM}{D_\mathrm{M}}
\newcommand{\rdrag}{r_\mathrm{drag}}
\newcommand{\zeff}{z_\mathrm{eff}}

\newcommand\aver[1]{\left\langle#1\right\rangle}
\newcommand\Leg[1]{\mathcal{L}_{#1}}

\newcommand\vr{\vect{r}}

\newcommand{\highlight}{}

\title[Reducing Variance of RSD Measurements]{Reducing the Variance of Redshift Space Distortion Measurements from Mock Galaxy Catalogues with Different Lines of Sight}

\author[A. Smith et al.]{\parbox{\textwidth}{
Alex Smith,$^{1}$\thanks{E-mail: alexander.smith@cea.fr}
Arnaud de Mattia,$^{1}$
Etienne Burtin,$^{1}$
Chia-Hsun Chuang,$^{2}$
Cheng Zhao$^{3}$
 } \vspace*{4pt} \\ 
$^{1}$ IRFU, CEA, Universit\'e Paris-Saclay, F-91191 Gif-sur-Yvette, France\vspace*{-2pt}\\
$^{2}$ Kavli Institute for Particle Astrophysics and Cosmology, Stanford University, 452 Lomita Mall, Stanford, CA 94305, USA\vspace*{-2pt} \\
$^{3}$ Institute of Physics, Laboratory of Astrophysics, \'Ecole Polytechnique F\'ed\'erale de Lausanne (EPFL), Observatoire de Sauverny, 1290 Versoix, Switzerland\vspace*{-2pt} 
}

\date{Accepted XXX. Received YYY; in original form ZZZ}

\pubyear{2020}

\begin{document}
\label{firstpage}
\pagerange{\pageref{firstpage}--\pageref{lastpage}}
\maketitle

\begin{abstract}
Accurate mock catalogues are essential for assessing systematics in the cosmological analysis of large galaxy surveys. Anisotropic two-point clustering measurements from the same simulation show some scatter for different lines of sight (LOS), but are on average equal, due to cosmic variance. This results in scatter in the measured cosmological parameters. We use the OuterRim N-body simulation halo catalogue to investigate this, considering the 3 simulation axes as LOS. The quadrupole of the 2-point statistics is particularly sensitive to changes in the LOS, with sub-percent level differences in the velocity distributions resulting in $\sim1.5\sigma$ shifts on large scales. Averaging over multiple LOS can reduce the impact of cosmic variance. We derive an expression for the Gaussian cross-correlation between the power spectrum multipole measurements, for any two LOS, including shot noise, and the corresponding reduction in variance in the average measurement. Quadrupole measurements are anti-correlated, and for three orthogonal LOS, the variance on the average measurement is reduced by more than $1/3$. We perform a Fisher analysis to predict the corresponding gain in precision on the cosmological parameter measurements, which we compare against a set of 300 extended Baryon Oscillation Spectroscopic Survey (eBOSS) emission line galaxy (ELG) EZmocks. The gain in $\fsig$, which measures the growth of structure, is also better than $1/3$. Averaging over multiple LOS in future mock challenges will allow the RSD models to be constrained with the same systematic error, with less than 3 times the CPU time.
\end{abstract}




\section{Introduction} 

Measurements of the spatial distribution of galaxies in large galaxy surveys allow us 
to probe the expansion history of the Universe and the growth of structure, 
testing the \LCDM\ cosmological paradigm and constraining models of dark energy.
In galaxy surveys, distances are inferred from redshifts, and each galaxy redshift
includes a component which is due to the peculiar velocity of the galaxy 
along the line of sight of the observer. This results in redshift space 
distortions (RSD), where structures appear flattened on large scales,
due to the coherent infall of galaxies to overdense regions \citep{Kaiser1987}, while on small scales, 
the random virial motion of galaxies results in elongated Fingers-of-God \citep{Jackson1972}. 

For a sample of galaxies at some effective redshift $\zeff$, the redshift-space
clustering statistics can be used to measure the growth of structure \citep{Guzzo2008}.
By fitting models of the redshift-space clustering to the data, the quantity 
$\fsig(\zeff)$, can be inferred, where $f=d\ln D(a) / d\ln a$ is the
linear growth rate, $D(a)$ is the linear growth function, and $\sigma_8$ 
is the rms of the density field in spheres of radius $8~\hMpc$.
In general relativity, the growth rate is related to the matter density through 
$f \approx \Omega_\mathrm{m}^{0.55}$ \citep[e.g.][]{Linder2005}.
Therefore, RSD measurements provide
a test for general relativity.

In addition to RSD, galaxy surveys can be used to measure baryon acoustic oscillations 
\citep[BAO; e.g.][]{Cole2005,Eisenstein2005}. 
Acoustic waves in the early Universe lead to an enhancement in the
present day galaxy distribution at a characteristic BAO scale of $\sim 100~\hMpc$. 
Therefore, BAO measurements from galaxy samples at different effective redshifts
can be used as a standard ruler to trace out the expansion history of the Universe.
If an incorrect fiducial cosmology is assumed when calculating cosmological distances
from redshifts, the BAO position is scaled by the parameters $\apar$ and $\aperp$, parallel 
and perpendicular to the line of sight, due to the Alcock-Paczynski effect \citep{Alcock1979}.
Measurements of $\apar$ and $\aperp$ can be used to determine the transverse comoving distance
$\dM(\zeff)/\rdrag$, and the Hubble distance $\dH(\zeff)/\rdrag$, 
where $\rdrag$ is the sound horizon at the drag epoch. The Hubble distance is related to the
Hubble parameter, $H(z)$, through $\dH(z)=c/H(z)$.

To validate the models that are used in the cosmological analysis of the survey data, it 
is important to use accurate mock galaxy catalogues. Since the `true' cosmology of the
mock is known, differences between the measured and `true' values of $\fsig$, $\dM/\rdrag$
and $\dH/\rdrag$ can be used to estimate the systematic uncertainties in the measurements
due to the modelling of galaxy clustering.

Typically, large N-body simulations are used, which have accurate clustering 
statistics down to small, non-linear scales. For the extended Baryon Oscillation 
Spectroscopic Survey \citep[eBOSS;][]{Dawson2016}, mock challenges have been
performed utilizing the OuterRim simulation \citep{Heitmann2019a}. The eBOSS
survey, which is part of SDSS-IV \citep{Blanton2017}, targeted 
luminous red galaxies (LRGs, $0.6<z<1.0$), emission line galaxies 
(ELGs, $0.6<z<1.1$) and quasars ($0.8<z<2.2$) as direct tracers of the matter field. 
For each of these tracers, mocks have been produced using the OuterRim box, 
which has a comoving side length of $3~\hGpc$.
Despite the large volume of the simulation, variations are seen in the
cosmological measurements when different lines of sight are chosen \citep{Alam2020,Smith2020}. 
Since the Universe is isotropic and homogeneous, all lines of sight
are equally valid, and there is no choice of observer position that is `better'
than any other. However, the finite box size of the simulation leads to
variations in the clustering statistics, due to cosmic variance. 
While the choice of line of sight has a very small impact on measurements
of $\apar$ and $\aperp$, offsets in the measured $\fsig$ for certain lines of sight 
are a few percent for the ELG mocks \citep{Alam2020}, 
and as large as $\sim 5\%$ for the quasar mocks \citep{Smith2020}, which is
at a level of more than $3\sigma$ of the modelling systematic uncertainty.\footnote{For the mocks in non-blind 
cosmologies, which do not include observational effects, the modelling systematic in $\fsig$ is
at a level of $\sim 1.5\%$ \citep{Smith2020}.}
Future surveys, such as the Dark Energy Spectroscopic Instrument \citep[DESI;][]{DESI2016a,DESI2016b},
aim to measure $\fsig$ to sub-percent precision.
In the mock challenges for eBOSS, these differences were mitigated by averaging over different lines of sight. 
The effect can also be mitigated by introducing a new estimator that averages over 
many lines of sight, such as the spherically averaged power spectrum \citep{Percival2009}.
When measuring $\fsig$ to the very high precision required in future surveys, it is important to 
understand this effect.

This paper is a companion paper to the eBOSS mock challenges that were performed to assess
the modelling systematics in the analysis of each of the different tracers. 
The LRG mock challenge is presented
in \citet{Rossi2020}, the ELG mock challenge is found in \citet{Alam2020}, and the
mock challenge for the quasar clustering sample is described in \citet{Smith2020}.
The final cosmological interpretation of the results from all the eBOSS analyses 
is presented in \citet{eBOSS_Cosmology2020}.

The aim of this paper is to understand why $\fsig$ measurements are
so sensitive to the choice of line of sight, and to calculate the improvement
in the measurements when averaging together multiple lines of sight.
This paper is outlined as follows: We give an overview of two-point clustering
statistics in Section~\ref{sec:two_point_clustering}, and the simulations we use in
Section~\ref{sec:simulations}.
Section~\ref{sec:clustering_statistics_los} investigates how the clustering
statistics are affected by the velocities in the simulation.
In Section~\ref{sec:cross_covariance_theory}, we derive an expression for the 
cross-correlation of power spectrum multipole measurements for two lines of sight,
and the corresponding reduction in the variance when averaging together multiple
measurements.
In Section~\ref{sec:fisher_analysis}, we perform a Fisher analysis to quantify
the reduction in the uncertainties of the measured cosmological parameters.
Finally, the conclusions are summarized in Section~\ref{sec:conclusions}.


\section{Two-point clustering} 
\label{sec:two_point_clustering}

\subsection{Correlation function}

The two-point correlation function of galaxies, $\xi(\vect{r})$, provides a measurement of the probability 
of finding a pair of galaxies with separation $\vect{r}$. It is defined as
\begin{equation}
\xi(\vect{r}) = \langle \delta(\vect{x}) \delta(\vect{x}+\vect{r}) \rangle,
\end{equation}
where $\delta(\vect{x})=\rho(\vect{x})/\bar{\rho}-1$ is the density contrast at position $\vect{x}$, $\bar{\rho}$
is the mean density, and the angled brackets indicate an ensemble average.

In redshift space, the correlation function is anisotropic, and we can measure the 2D correlation function
$\xi(s,\mu)$, where $s$ is the separation in redshift space, and $\mu$
is the cosine of the angle between the line of sight and pair separation vector. 
The information in the 2D correlation function can be expressed 
by decomposing $\xi(s,\mu)$ into Legendre multipoles,
\begin{equation}
\xi(s,\mu) = \sum_\ell \xi_\ell (s) \mathcal{L}_\ell(\mu),
\end{equation}
where $\mathcal{L}_\ell(\mu)$ is the $\ell^\mathrm{th}$ order Legendre polynomial.
The multipoles $\xi_{\ell}(s)$ can be calculated using
\begin{equation}
\xi_\ell(s) = \frac{2\ell + 1}{2} \int_{-1}^{1} \xi(s,\mu) \mathcal{L}_\ell(\mu) d\mu.
\label{eq:xi_multipoles}
\end{equation}
Only the first three even multipoles are non-zero in linear theory, which we call the monopole, $\xi_0(s)$,
quadrupole, $\xi_2(s)$, and hexadecapole, $\xi_4(s)$.

To compute the correlation function, the estimator of \citet{Landy1993} is commonly used,
\begin{equation}
\xi(s,\mu) = \frac{DD(s,\mu) - 2DR(s,\mu) + RR(s,\mu)}{RR(s,\mu)},
\end{equation}
where $DD(s,\mu)$, $DR(s,\mu)$ and $RR(s,\mu)$ are the normalized data-data, data-random
and random-random pair counts. For a periodic box, the $RR(s,\mu)$ counts can be
computed analytically, and only the $DD(s,\mu)$ pair counts are needed.

\subsection{Power spectrum}

Alternatively, the two-point clustering statistics can be expressed in Fourier space using 
the power spectrum, $P(k)$, defined as
\begin{equation}
\aver{\delta(\vect{k})\delta(\vect{k}^{\prime})} = \left(2\pi\right)^{3}\delta^D(\vect{k}+\vect{k}^{\prime})P(k),
\label{eq:Pk_definition}
\end{equation}
and is the Fourier transform of the correlation function. As with the correlation
function, the power spectrum can be decomposed into Legendre polynomials. For a periodic
box, this is given by
\begin{equation}
P_{\ell}(k) = \left(2\ell+1\right) \int \frac{d\Omega_{k}}{4\pi V} \delta_{g}(\vect{k})\delta_{g}(-\vect{k})\mathcal{L}_\ell(\hat{\vect{k}} \cdot \hat{\vect{\eta}}) - P_{\ell}^{\mathrm{noise}}(k),
\label{eq:power_spectrum_estimator_box}
\end{equation}
where $\delta_g(\vect{k})$ is the overdensity of galaxies in Fourier space, $\hat{\vect{\eta}}$ is
the unit line-of-sight vector, $V$ the volume of the box and $d\Omega_{k}$ is the solid angle.
The shot noise term, which is non-zero only for the monopole, is
\begin{equation}
P_{0}^{\mathrm{noise}} = \frac{1}{\bar{n}_g}
\end{equation}
where $\bar{n}_g$ is the mean number density of galaxies.


\section{Simulations} 
\label{sec:simulations}

\subsection{OuterRim simulation}
\label{sec:outerrim}

In this work, we use the halo catalogue from the OuterRim simulation \citep{Habib2016,Heitmann2019a,Heitmann2019b}.
The OuterRim simulation is a dark-matter-only N-body simulation, which uses a flat $\Lambda$CDM cosmology
that is consistent with the WMAP7 measurements \citep{Komatsu2011}, with $\Omega_\mathrm{cdm} h^2=0.1109$,
$\Omega_\mathrm{b} h^2=0.02258$, $h=0.71$, $\sigma_8=0.8$ and $n_s=0.963$. The simulation
contains 10,240$^3$ particles of mass $m_p=1.85\times 10^9~\hMsun$ within a box of comoving 
side length $3000~\hMpc$. Dark matter haloes are identified with a friends-of-friends (FOF) 
algorithm \citep{Davis1985}, with a linking length $b=0.168$.
We use the simulation snapshot at $z=1.433$ with a mass cut $M > 3 \times 10^{12} \hMsun$ applied,
giving a total number density of haloes of $4.9\times 10^{-4} (\hMpc)^{-3}$. 

This halo catalogue is comparable in redshift and linear bias to the eBOSS quasar sample.
We use the halo catalogue directly, and not the quasar mocks produced in \citet{Smith2020}, since the 
number density of quasars is low due to the $\sim 1\%$ duty cycle.

\subsection{EZmocks}
\label{sec:ezmocks}

We also utilize a set of 300 Effective Zel'dovich approximation Mocks \citep[EZmocks;][]{Chuang2015}
which were constructed for the eBOSS ELG sample \citep{Zhao2020}.
We use these mocks to quantify the uncertainties on cosmological parameter measurements for
a realistic galaxy survey, and the large number of mocks improves our statistics.
In each box, the density field is generated using the Zel'dovich approximation, and galaxies
are added using a parametrisation of the galaxy bias. The mocks are calibrated to the clustering
measurements of the eBOSS ELGs.

The EZmock boxes are in a Planck cosmology \citep{Planck2014} with
$\Omega_\mathrm{m} = 0.307115$, $\Omega_\mathrm{b} = 0.048206$, $h=0.6777$,
$\sigma_8=0.8225$ and $n_s=0.9611$, and have a box size of $5000~\hMpc$. 
We use boxes at $z=0.876$, with a number density of $6.4 \times 10^{-4}~(\hMpc)^{-3}$.


\section{Clustering statistics for different lines of sight} 
\label{sec:clustering_statistics_los}

\subsection{Clustering measurements}

Transforming the coordinates of a tracer from its real-space to redshift-space
position requires a choice to be made for the observer position. For a periodic
box, it is common to use the plane parallel approximation, where the observer is 
positioned at infinity along e.g. the $x$-axis of the simulation. The halo positions
are then displaced using the $x$-component of the velocity.
In this case, the comoving position in redshift space is given by
\begin{equation}
\vect{s} = \vect{r} + \frac{v_x}{aH(z)} \vect{\hat{x}},
\label{eq:rsd_transform}
\end{equation}
where $\vect{\hat{x}}$ is a unit vector parallel to the $x$-axis, 
and $v_x$ is the $x$-component of proper velocity.
In the case of a periodic box, periodic boundary conditions are then applied.

In real space, the clustering measurements are not impacted by the choice of observer 
position, since the distance between a pair of galaxies is unaffected by the line of sight.
In redshift space, this is not true. The velocity used to transform to redshift
space depends on the observer, so we expect to see some variation in the clustering 
measurements for different lines of sight.

\begin{figure*} 
\centering
\includegraphics[width=\linewidth]{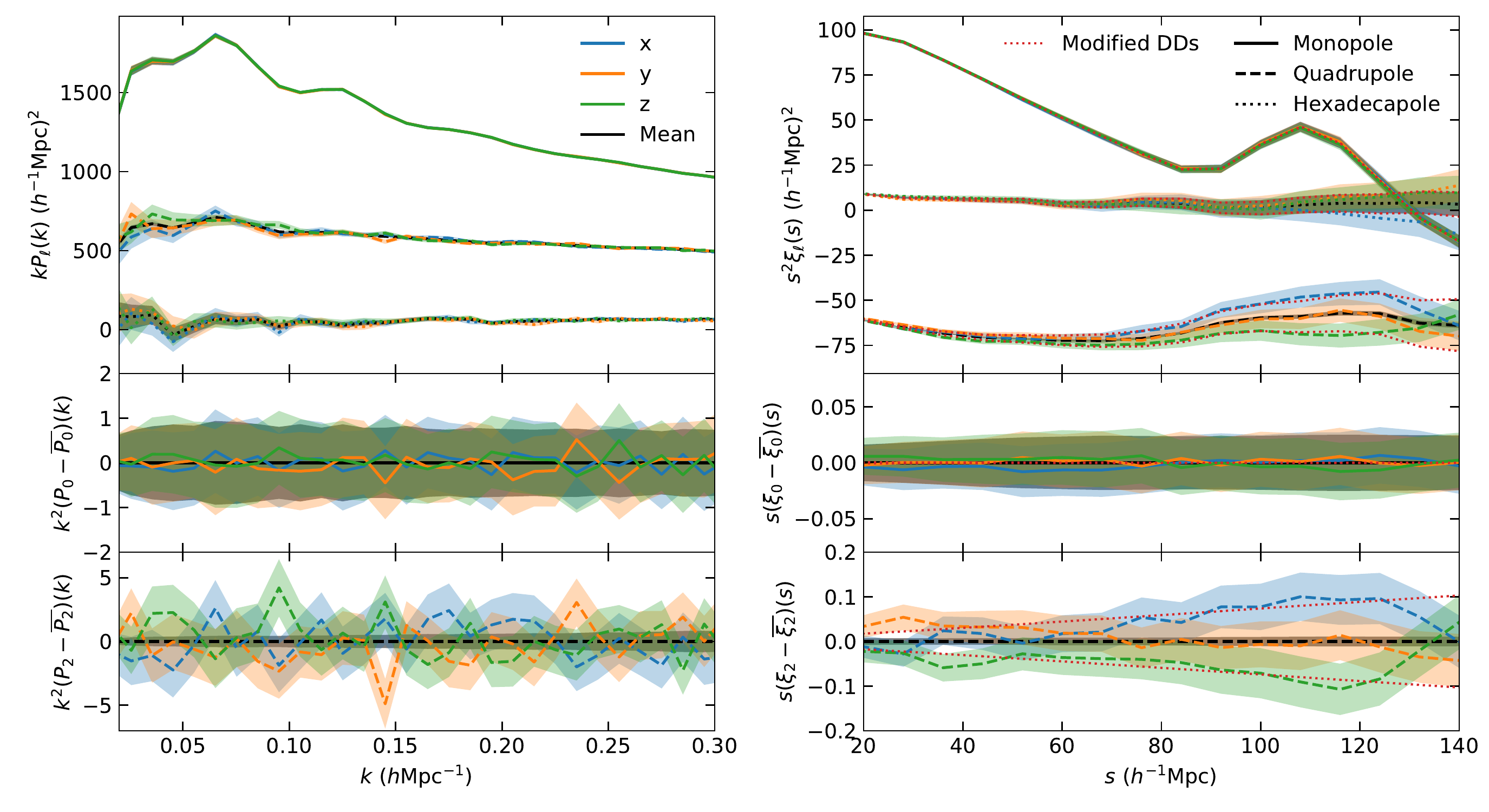}
\caption{\textit{Left panel}: Power spectrum multipole measurements from the OuterRim halo catalogue, with observer
at infinity in the $x$ (blue), $y$ (yellow) and $z$-direction (green). The average of these measurements is shown in black.
The monopole, quadrupole and hexadecapole are indicated by the solid, dashed and dotted lines, respectively.
Shaded regions indicate jackknife errors, estimated using 512 jackknife samples.
Lower panels show the difference between each measurement and the average measurement, for the 
monopole and quadrupole.
\textit{Right panel}: Same as the left panel, but for the correlation function multipoles, $\xi_\ell(s)$.
The red dotted curve shows the result of modifying the average DD pair counts by a factor of 
$7 \times 10^{-4}$, for bins with $\mu>0.75$ (see Section~\ref{sec:pair_counts}).}
\label{fig:pk_xi}
\end{figure*}

This is illustrated in Fig.~\ref{fig:pk_xi}, where we show the power spectrum and 
correlation function multipoles of the OuterRim halo catalogue
when transforming to redshift space with an observer at infinity in the $x$, $y$ and 
$z$-direction. The power spectrum is measured using 
the algorithm from \textsc{nbodykit} \citep{Hand2018}, while the correlation function is computed 
using the publicly available parallelized code \textsc{twopcf}.\footnote{\url{https://github.com/lstothert/two_pcf}}
\highlight{The uncertainties in the measurements, (shown by the shaded regions) are estimated using 
the jackknife resampling technique, where the box has been split into 512 jackknife subsamples 
(each with size $375~\hMpc$).}
For the power spectrum, the scatter between the monopole measurements is very
small, at a level of $\sim 0.2\%$, with larger variations in the quadrupole, at a level of $\sim 2\%$.
For the correlation function, the differences between the measurements for different observers show 
much more scale dependency, with larger variations on large scales. At a scale of $100~\hMpc$, the
scatter in the monopole is at a level of $\sim 2\%$, and the scatter in the quadrupole
is as large as $\sim 10\%$, with smaller differences on small scales.

While the variations in the quadrupole measurements are surprisingly 
large, this scatter is expected due to cosmic variance 
(see Section~\ref{sec:cross_covariance_theory}). Since measurements of $\fsig$
are proportional to the amplitude of the quadrupole, a large offset in the
quadrupole will also lead to large offsets in measurements of $\fsig$. It is therefore important
to understand these large variations in the quadrupole measurements
for mock challenges.

The halo catalogue from OuterRim that we are using is comparable to the
eBOSS quasar sample. The variations in the clustering measurements for different 
observer positions is large for the quasars, but note
that the results in the following subsections are not specific to any
single tracer.

\subsection{Pair counts}
\label{sec:pair_counts}

\begin{figure} 
\centering
\includegraphics[width=\linewidth]{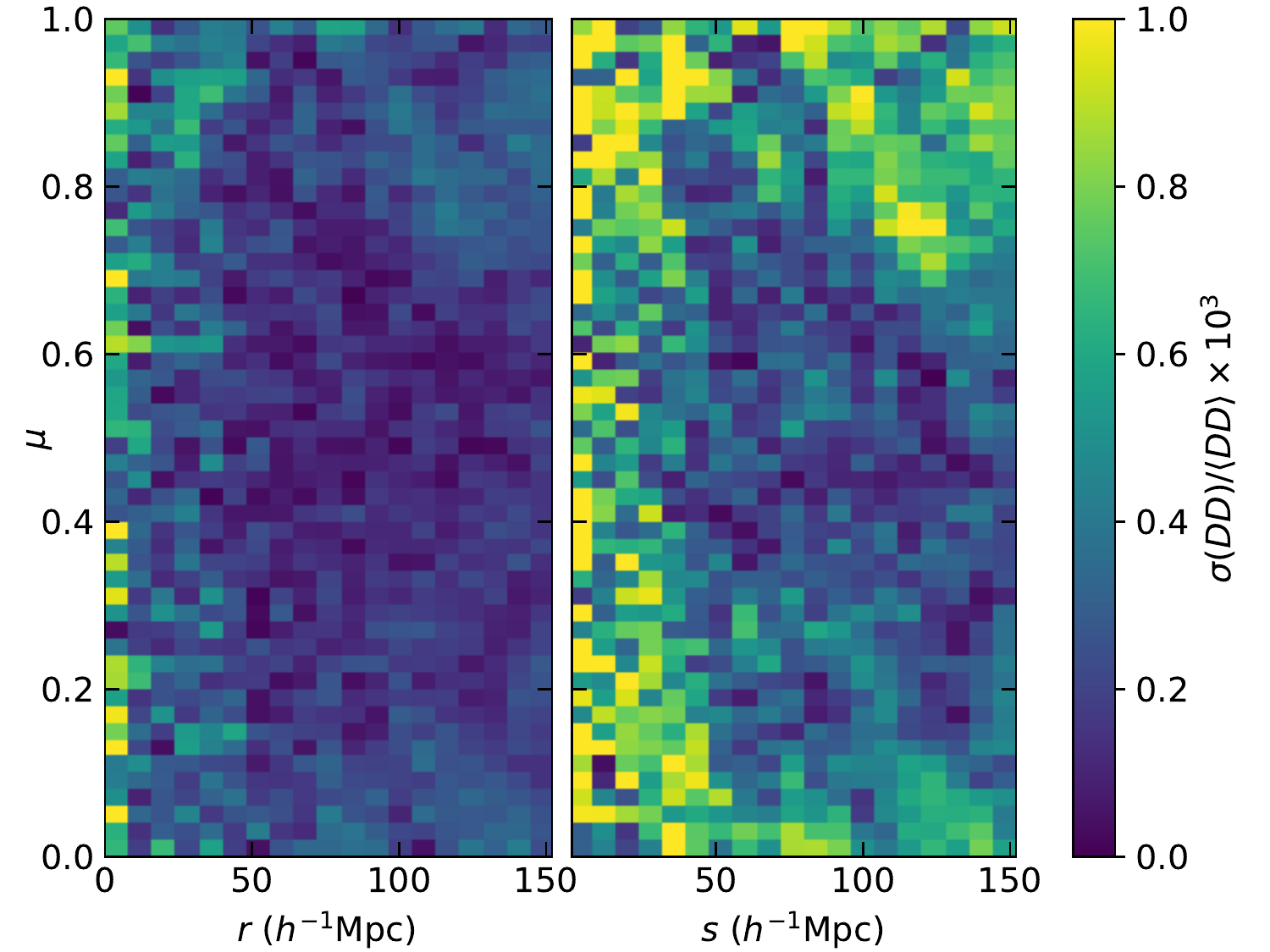}
\caption{Rms of the $DD(r,\mu)$ counts, divided by the mean, from the three
measurements of $DD(r,\mu)$ with observer at infinity in the direction of
each of the simulation axes. 
\textit{Left}: in real space. \textit{Right}: in redshift space.}
\label{fig:dd_haloes}
\end{figure}

To understand how the correlation function is affected by the observer position, it is more 
convenient to first consider the $DD$ pair counts, rather than the correlation function
directly.
The variation in the $DD$ counts from the OuterRim halo catalogue, 
measured by an observer at infinity along each of the three
simulation axes, is shown in Fig.~\ref{fig:dd_haloes}. 
The left panel is in real space, where the colour indicates the standard deviation of the three measurements
of $DD(r,\mu)$, measured in bins of $r$ and $\mu$, divided by the mean. 
In real space, the variation between the three measurements is very small. The separation
between each pair does not change, and the variation that is seen is due to pairs being placed in different $\mu$
bins when observed from different directions. When integrating over $\mu$, the total number
of pairs in each bin of $r$ is constant.

The right hand panel of Fig.~\ref{fig:dd_haloes} shows the variation in the $DD$ counts in 
redshift space. For pairs with large separations, the difference in the $DD(s,\mu)$ counts is 
close to zero for $\mu \lesssim 0.75$, but for $\mu \gtrsim 0.75$, the scatter is of the order of 
$\sim 10^{-3}$. The redshift-space separation, $s$, between a pair of galaxies will be
different for different observers, due to the different velocity components along each line of sight.
Pairs separated along the line of sight (with $\mu \sim 1$) are more strongly affected
by velocities, since the separation and velocity vectors are parallel. This variation
in $DD(s,\mu)$ leads to
the differences in the clustering seen in Fig.~\ref{fig:pk_xi}. Pairs with
$\mu \sim 1$ are weighted more strongly when calculating the quadrupole 
(in Eq.~\ref{eq:xi_multipoles}, the Legendre polynomial $\Leg{2}(\mu)=1$ when $\mu=1$), 
explaining why the differences in the quadrupole are much larger than for the monopole.

It is important to note that the differences in the $DD$ counts are very small (less than $10^{-3}$), yet
this is enough to shift the amplitude of the quadrupole on large scales by as much as $\sim 10 \%$.
This is illustrated by the red dotted curves in
Fig.~\ref{fig:pk_xi}, where the average correlation function (black curve) is modified 
so that the $DD(s,\mu)$ counts are increased (and decreased) by 0.07\% for $\mu > 0.75$ 
(and the $\mu<0.75$ counts are modified by a very small factor
to keep the total number of $DD$ pairs unchanged). 
This modification is able to reproduce what is measured in the
halo catalogue. 

\begin{figure} 
\centering
\includegraphics[width=\linewidth]{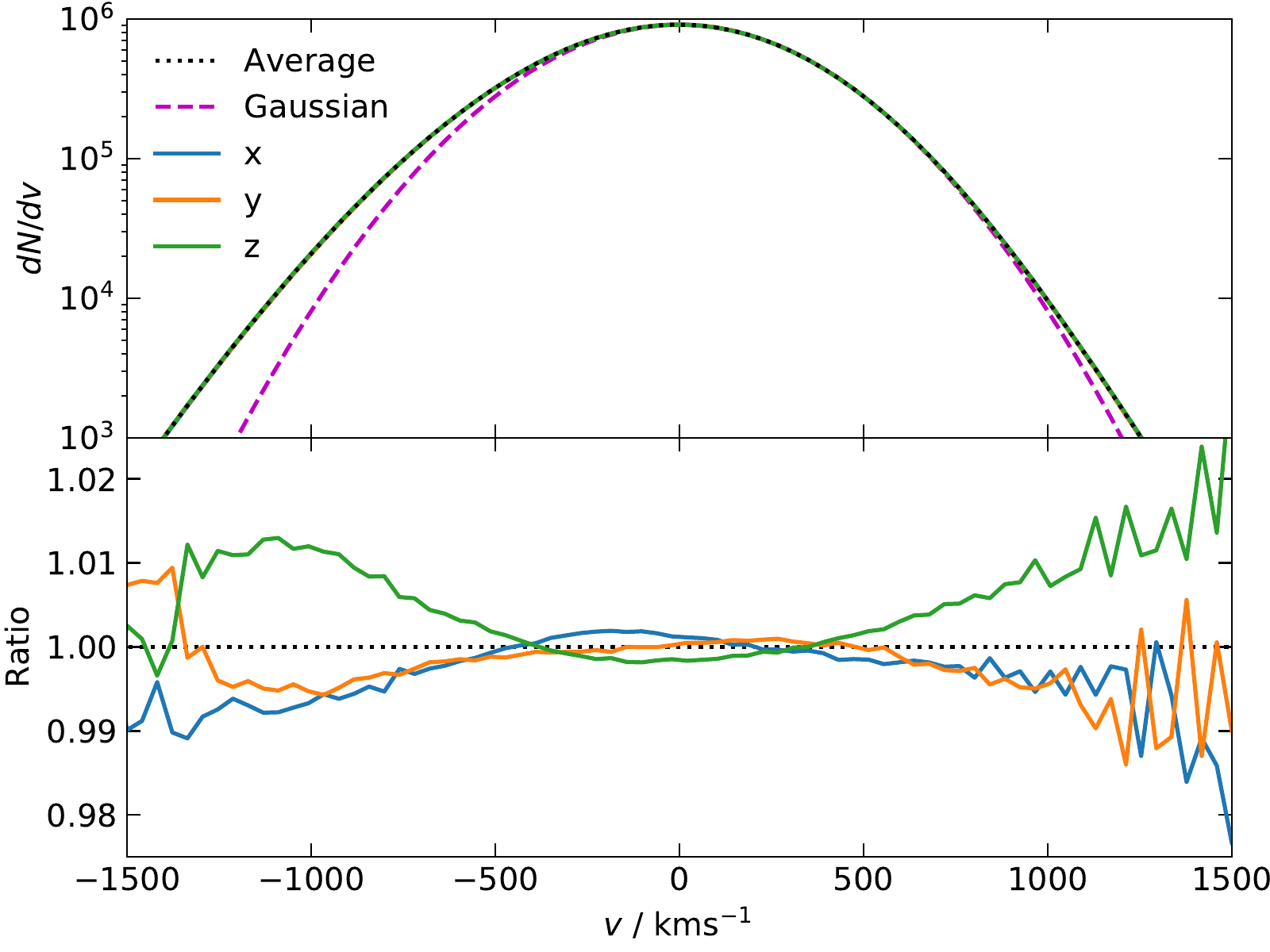}
\caption{\textit{Top panel}: Distribution of the $x$, $y$ and $z$ components of pairwise halo velocities
(coloured curves), for pairs of haloes with separation $99 < r < 101~\hMpc$. The dotted
black curve is the average distribution. \highlight{The dashed magenta curve shows a Gaussian, which
highlights the non-Gaussianity in the tails of the velocity distribution.}
\textit{Bottom panel}: Ratio to the average distribution.}
\label{fig:halo_velocity_distribution}
\end{figure}

The pairwise velocity distributions, which are responsible for the variations in the $DD$ counts,
are shown in Fig.~\ref{fig:halo_velocity_distribution}.
The upper panel shows the $x$, $y$ and $z$ components, for pairs with separation 
$99 < r < 101~\hMpc$. The pairwise velocity, $v_\mathrm{pair,x}$,
in the $x$-direction is defined as the difference between the $x$-components of the
individual velocities, $v_\mathrm{pair,x} = v_{1,x} - v_{2,x}$, where the $x$-position $x_1 > x_2$. 
This distribution looks Gaussian, but is slightly skewed towards negative
velocities (i.e. infalling pairs), as expected, \citep[e.g.][]{Juszkiewicz1998,Tinker2007,Bianchi2015,Kuruvilla2018}
and the distribution looks almost identical for the three directions.
The skewness of the distribution becomes larger on smaller scales, where
both members of the pair reside in the same dense environment.
However, the bottom panel of Fig.~\ref{fig:halo_velocity_distribution}
shows the ratio to the average distribution, revealing that there are small differences, which 
are mostly seen in the variance of the distribution.

To summarise, for an observer at infinity, differences are seen in the measurements of the 
quadrupole and hexadecapole for different observer positions. This effect is due to the finite
box size of the simulation, and is explained by small
differences in the pairwise velocity distributions projected along the three simulation axes.
As the volume is increased towards infinity, the velocity
distributions will converge, and the effect will diminish. But it is important to note that even 
for a simulation as large as OuterRim, with a $3~\hGpc$ box, the choice of observer
position has a significant impact on measurements of the quadrupole, which will propagate
to the measurements of $\fsig$.

\subsection{Velocities}
\label{sec:velocities}

In this subsection, we study in more detail the velocity distributions, which
are responsible for the variations seen in the clustering measurements.
Understanding the velocities is important, as it is one of the components needed in the 
streaming model \citep[e.g.][]{Fisher1995,Scoccimarro2004}, which is often used to predict the redshift-space
clustering of galaxies from a model of the real-space clustering \citep[e.g.][]{Reid2011,Wang2014}.

\begin{figure} 
\centering
\includegraphics[width=\linewidth]{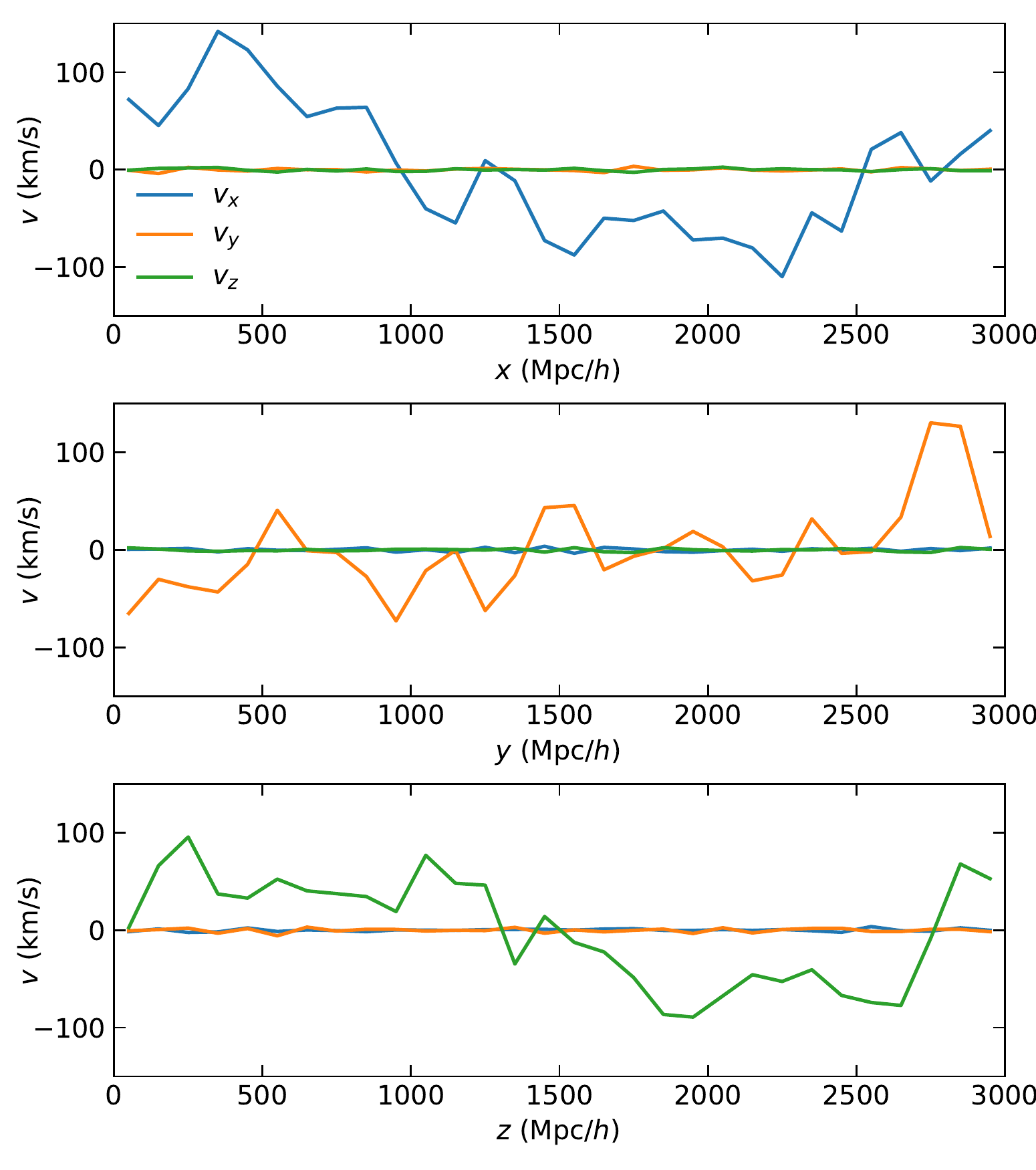}
\caption{Average of the velocity components $v_x$ (blue), $v_y$ (yellow) and $v_z$ (green),
in bins of $x$ (upper panel), $y$ (middle panel) and $z$ (lower panel).}
\label{fig:outerrim_velocities}
\end{figure}

Fig.~\ref{fig:outerrim_velocities} shows the distribution of the individual velocities of each object
in the box. The average of each of the three velocity components, $v_i$, is shown in bins 
of position along each axis, $j$. 
When $i=j$, (i.e. the velocity component is parallel with the axis in which it is binned)
variations can been seen, which is due to the coherent motion of haloes within
the large-scale structure of the finite box. 
When $i \neq j$, it is interesting to note that the average velocity is zero. 
\highlight{This is because of the longitudinal modes in the simulation, which dominate on large scales.
While there is some vorticity, this is very weak, and only becomes important on small scales
\citep[e.g.][]{Pueblas2009}.}
Another way to think about this is to consider a single spherically symmetric overdense region, 
which is surrounded by infalling haloes. The average
$v_x$ of the haloes, in bins of $x$, will be positive for small $x$, and negative at
large $x$, since the average motion of the haloes is towards the centre of the overdense region. 
However, the average $v_y$ and $v_z$, in bins of $x$, is zero, due to the spherical
symmetry. 

\begin{figure} 
\centering
\includegraphics[width=\linewidth]{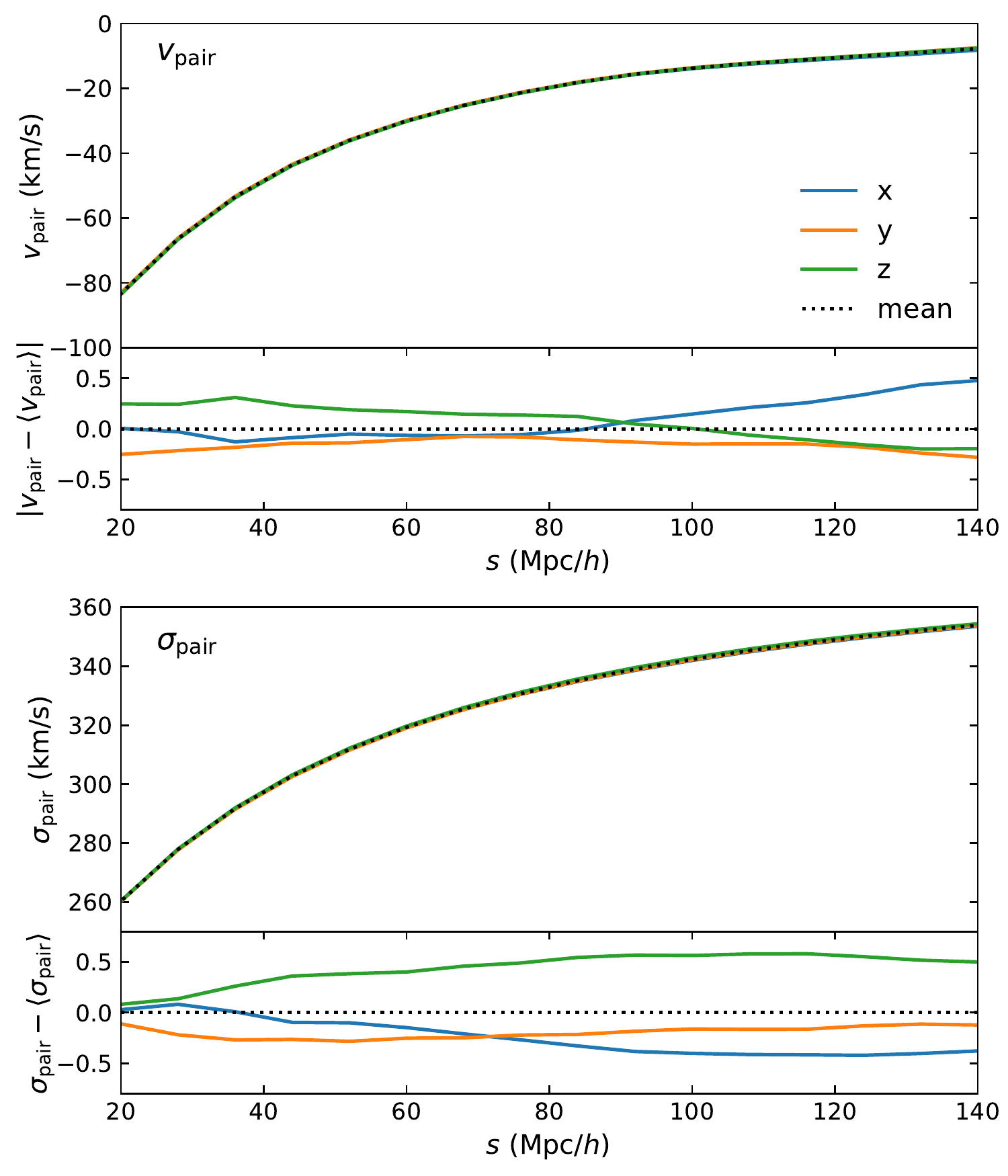}
\caption{\textit{Top:} mean of the pairwise velocity distribution for the $x$ (blue), $y$ (yellow)
and $z$ (green) components of velocity. The black dotted curve is the average of these three measurements.
The lower panel shows the difference to the average measurement.
\textit{Bottom:} As above, but for the rms of the pairwise velocity distribution.}
\label{fig:infall_velocities}
\end{figure}

However, it is the pairwise velocities, and not the individual velocities, that are responsible
for the variation seen in the clustering measurements. The distribution of pairwise velocities
for pairs in a single bin of separation (at $100~\hMpc$) is shown in Fig.~\ref{fig:halo_velocity_distribution}.
The mean and standard deviation of the same distribution is shown in Fig.~\ref{fig:infall_velocities},
but as a function of separation. The top panel shows the $x$, $y$ and $z$-components of the 
mean pairwise velocity, $v_\mathrm{pair}$, and the difference to the average $v_\mathrm{pair}$. 
Negative velocities indicate that
the pairs are infalling, which is expected, due to the coherent infall of haloes towards 
dense regions. The magnitude of the infall velocity increases with decreasing pair separation, 
and this is consistent with similar measurements from other simulations, and the
predictions from perturbation theory models \citep[e.g.][]{Cuesta2020}. On very small
scales, the average velocity should tend towards zero, due to the random motion within virialized
structures, but we don't see this, since our catalogue contains massive haloes, and we only show
scales above $20~\hMpc$.

The bottom panel of Fig.~\ref{fig:infall_velocities} shows the rms of the velocity distribution,
$\sigma_\mathrm{pair}$, as a function of pair separation for each line of sight, and the difference
from the mean value of $\sigma_\mathrm{pair}$. As the pair separation is reduced, the rms of
the velocity distribution also decreases. It is interesting to note that the differences in 
$\sigma_\mathrm{pair}$ mirror what is seen in the correlation function quadrupole,
from Fig.~\ref{fig:pk_xi}.
E.g., $\sigma_{\mathrm{pair},z}$ is largest, and the quadrupole measurement with the 
$z$-direction as the line of sight also has the highest amplitude (i.e. is more negative). 
This suggests that it is the rms of the pairwise velocity distribution that is responsible for
the differences seen in the quadrupole measurements.

\begin{figure} 
\centering
\includegraphics[width=0.75\linewidth]{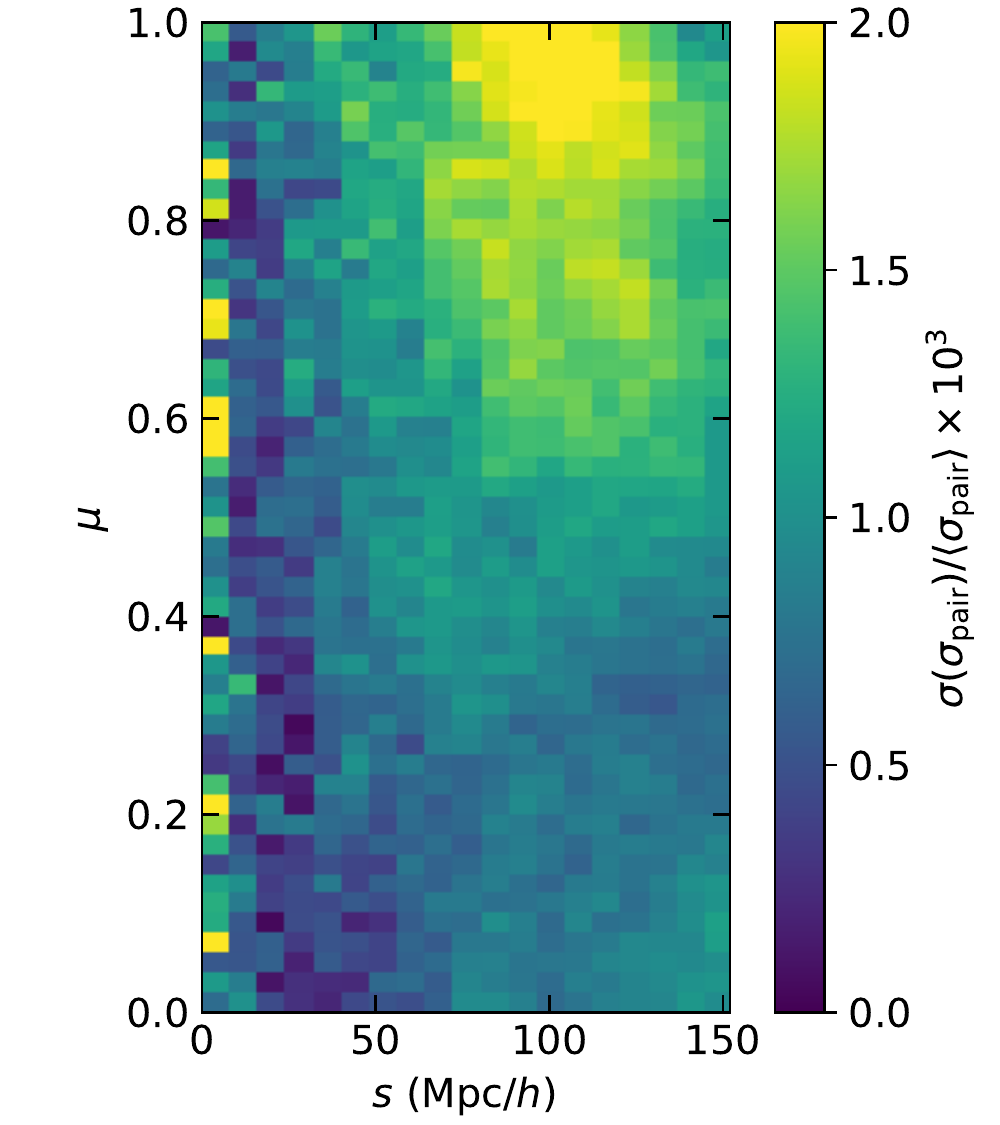}
\caption{Scatter in the rms of the pairwise velocity distributions, $\sigma_\mathrm{pair}$, in bins 
of $s$ and $\mu$, for the three observer positions. The colour indicates the rms of the 
three $\sigma_\mathrm{pair}$ measurements, divided by the mean, scaled by $10^3$.}
\label{fig:sigma_scatter}
\end{figure}

Fig.~\ref{fig:sigma_scatter} shows the standard deviation in the three velocity rms measurements,
divided by the mean, in bins of $s$ and $\mu$, similar to Fig.~\ref{fig:dd_haloes} for the $DD$ pair counts.
The variations in the rms are larger than the variations in the $DD$ counts, but they are the same
order of magnitude, and the largest variations are similarly seen in large $s$ bins where 
$\mu$ is close to 1. While the measurements of the quadrupole depend on the original positions of
the haloes, and the mean and the rms of the velocity distributions, this shows that it is the
rms of the velocity distribution that has the greatest contribution to the relative amplitudes
of the quadrupole. 

\subsection{Gain in uncertainties} \label{sec:gain_uncertainties_measured}

\highlight{The uncertainties in Fig.~\ref{fig:pk_xi} were estimated using the jackknife resampling technique.}
When averaging together the three measurements that use different lines of sight, the uncertainties 
in the combined monopole measurement is comparable to the three individual measurements.
However, for the quadrupole, the uncertainties in the combined measurement are greatly reduced, 
in both the correlation function and power spectrum. 

\begin{figure*} 
\centering
\includegraphics[width=\linewidth]{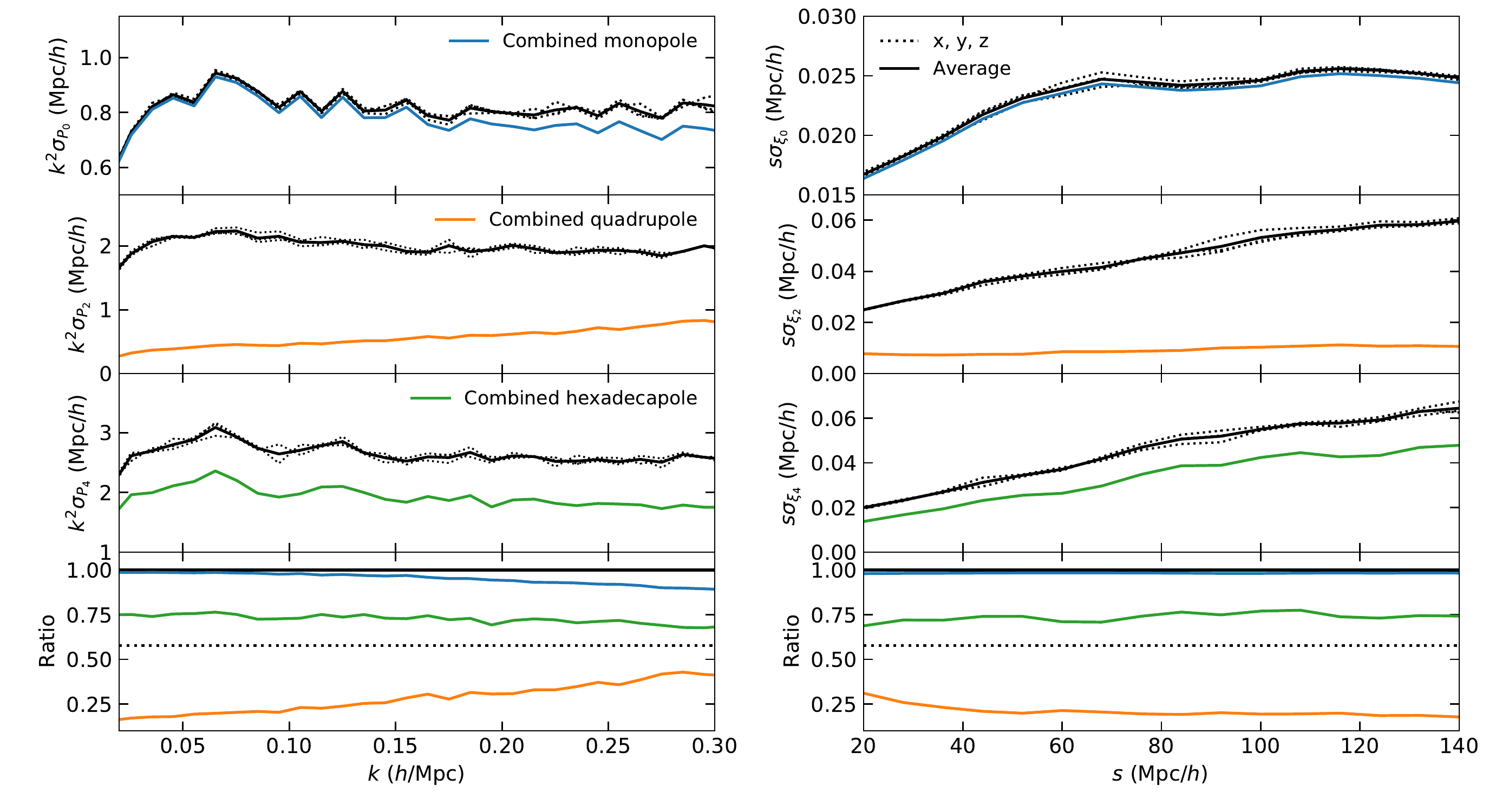}
\caption{\textit{Left}: Errors in the power spectrum multipole measurements from OuterRim, estimated
using 512 jackknife subsamples. The first 3 panels show the errors in the monopole, quadrupole and
hexadecapole, respectively. Black dotted curves show the errors for the 3 different lines of sight, 
and the black solid curve is the average of these 3. The coloured curves show the errors when the
3 lines of sight are combined. The ratios are shown in the lower panel, indicating the reduction in
errors when combining the lines of sight. The horizontal lines
indicate ratios of 1 and $1/\sqrt{3}$.
\textit{Right}: same as the left panels, but for the correlation function measurements.}
\label{fig:pk_xi_err}
\end{figure*}

This reduction in the uncertainties is shown more clearly in Fig.~\ref{fig:pk_xi_err}.
The upper three sets of panels show the jackknife uncertainties in the power spectrum (left)
and correlation function (right), for the three individual measurements (black dotted curves),
the mean of the three uncertainties (black solid curve), and the uncertainty in the
combined measurement (coloured curves). The ratio is shown in the lower panels. 

On large scales (i.e. small $k$), there is very little change in the uncertainties in the
power spectrum monopole after combining the three measurements, and the ratio is very close
to 1. This is not surprising, since the three measurements of the monopole are highly correlated,
so very little information is gained when combining them. On small scales (large $k$), there
is a small reduction in the uncertainties, while in the correlation function, the ratio is consistent with 1
over all scales. There is a larger reduction in the uncertainties on the hexadecapole, by
a factor of $\sim 0.75$, in both the power spectrum and correlation function on large scales.

The most striking result is for the quadrupole, where the uncertainties are 
reduced by a factor $\sim 0.2$ on large scales. In the case that the measurements of a
quantity are completely uncorrelated, combining the three lines of sight would be 
equivalent to increasing the volume by a factor of 3, and therefore the uncertainties 
in the combined measurement would be reduced by a factor of $1/\sqrt{3} \approx 0.58$ 
(indicated by the horizontal dotted line in the lower panels of Fig.~\ref{fig:pk_xi_err}).
It is remarkable that the gain in the quadrupole measurement is much greater than this.

In the next section, we aim to understand this result. 
We calculate the \highlight{cross-correlation} between power spectrum measurements for two different lines of sight,
and show how this relates to the gain in the uncertainties when combining multiple measurements together.

\section{\highlight{Cross-correlation} between two different lines of sight}
\label{sec:cross_covariance_theory}

The measurements of galaxy clustering from a simulation are impacted by the
choice made for the line of sight. In this section we estimate the \highlight{cross-correlation}
for the power spectrum multipoles, measured with two lines of sight, $u$ and $v$. 
In Section~\ref{sec:cross_covariance_orthogonal}, we derive an expression for
the \highlight{cross-correlation} in the case where $u$ and $v$ are orthogonal, neglecting shot noise.
In Section~\ref{sec:cross_covariance_shot_noise}, we generalise this for any two
lines of sight, and include shot noise. We show how the \highlight{cross-correlation} is related to
the gain in the uncertainties in the power spectrum measurements in
Section~\ref{sec:gain_uncertainty_theory}. In Section~\ref{sec:discussion}, we provide a more
intuitive explanation of these results.

\subsection{Orthogonal lines of sight}
\label{sec:cross_covariance_orthogonal}

The cross-covariance matrix of the power spectrum multipoles, with different lines of sight,
can be written as
\begin{align}
C_{\ell\ell^{\prime}ij}^{uv} &= \Big\langle\big[\hat{P}_\ell^{u}(k_i) - \langle \hat{P}_\ell^{u}(k_i)\rangle \big]\big[ \hat{P}_{\ell^{\prime}}^{v}(k_j) - \langle \hat{P}_{\ell^{\prime}}^{v}(k_j)\rangle \big]\Big\rangle \\
&= C_{\ell\ell^{\prime}ij} \rho_{\ell\ell^{\prime}ij}^{uv},
\label{eq:cross-covariance}
\end{align}
where $C_{\ell\ell^{\prime}ij}$ is the covariance matrix of power spectrum measurements with the same line of sight, and
$\rho_{\ell\ell^{\prime}ij}^{uv}$ is the \highlight{cross-correlation}, which we aim to calculate. By rearranging 
Eq.~\ref{eq:cross-covariance}, we can write $\rho_{\ell\ell^{\prime}ij}^{uv}$ as
\begin{align}
\rho_{\ell\ell^{\prime}ij}^{uv} &= \frac{\left\langle\big[\hat{P}_\ell^{u}(k_i) - \langle \hat{P}_\ell^{u}(k_i)\rangle \big]\big[ \hat{P}_{\ell^{\prime}}^{v}(k_j) - \langle \hat{P}_{\ell^{\prime}}^{v}(k_j)\rangle \big]\right\rangle}{C_{\ell\ell^{\prime}ij}} \\
&= \frac{\big\langle \hat{P}^{u}_\ell(k_i) \hat{P}^{v}_{\ell^\prime}(k_j) \big\rangle - \big\langle \hat{P}^{u}_\ell(k_i)\big\rangle \big\langle \hat{P}^{v}_{\ell^\prime}(k_j)\big\rangle}{\big\langle \hat{P}^{u}_\ell(k_i)\hat{P}^{u}_{\ell^\prime}(k_j)\big\rangle - \big\langle \hat{P}^{u}_\ell(k_i)\big\rangle\big\langle \hat{P}^{u}_{\ell^\prime}(k_j)\big\rangle},
\label{eq:cross_correlation_P}
\end{align}
where we arbitrarily took $u$ as line of sight in the denominator.
The first term in the numerator of Eq.~\ref{eq:cross_correlation_P} can be written as
\begin{align}
\aver{\hat{P}_{\ell}^{u}(k)\hat{P}_{\ell^{\prime}}^{v}(k^{\prime})} \propto \int & d\Omega_{u} d\Omega_{v} \aver{\delta_{g}^{u}(\vect{k})\delta_{g}^{u}(-\vect{k})\delta_{g}^{v}(\vect{k}^{\prime})\delta_{g}^{v}(-\vect{k}^{\prime})} \nonumber \\
& \Leg{\ell}(\cos{\theta_{u}})\highlight{\Leg{\ell^{\prime}}}(\cos{\theta_{v}}), 
\end{align}
where the prefactors can be dropped, since they are the same in both the numerator and denominator of 
Eq.~\ref{eq:cross_correlation_P}, and will cancel out.
Assuming Gaussianity of $\delta_{g}$, Wick's theorem gives
\begin{align}
&\aver{\delta_{g}^{u}(\vect{k})\delta_{g}^{u}(-\vect{k})\delta_{g}^{v}(\vect{k}^{\prime})\delta_{g}^{v}(-\vect{k}^{\prime})} \nonumber \\
&= \aver{\delta_{g}^{u}(\vect{k})\delta_{g}^{u}(-\vect{k})}\aver{\delta_{g}^{v}(\vect{k}^{\prime})\delta_{g}^{v}(-\vect{k}^{\prime})} \nonumber \\
&+ \aver{\delta_{g}^{u}(\vect{k})\delta_{g}^{v}(\vect{k}^{\prime})}\aver{\delta_{g}^{u}(-\vect{k})\delta_{g}^{v}(-\vect{k}^{\prime})} \nonumber \\
&+ \aver{\delta_{g}^{u}(-\vect{k})\delta_{g}^{v}(\vect{k}^{\prime})}\aver{\delta_{g}^{u}(\vect{k})\delta_{g}^{v}(-\vect{k}^{\prime})}.
\end{align}
The first of these terms is $\big\langle \hat{P}_\ell^{u}(k_i)\big\rangle \big\langle \hat{P}_{\ell^{\prime}}^{v}(k_j)\big\rangle$, 
which can be dropped, since it is being subtracted by the second term in the numerator of
Eq.~\ref{eq:cross_correlation_P}.

\highlight{
Let us now compute the cross-correlation, $\aver{ \delta_{g}^{u}(\vect{k})\delta_{g}^{v}(\vect{k}^{\prime}) }$. 
The RSD displacement field along line of sight $u$ can be written as $\psi_{u}(\vr)$ such that the redshift-space 
position, $\vect{s}_{u}$, is related to the real-space position, $\vr$, through 
$\vect{s}_{u} = \vr + \psi_{u}(\vr) \vect{\hat{u}}$ (see Eq.~\ref{eq:rsd_transform}). 
It is then straighforward to extend the redshift-space power spectrum \citep[equation~4 of][]{Taruya2010:1006.0699v1} to the cross power spectrum of the redshift-space density contrast, seen with two lines of sight $u$ and $v$,
\begin{align}
P_{s}^{uv}(\vect{k}) &= \int d^{3} \vect{x} e^{- i \vect{k} \cdot \vect{x}} \left\langle e^{i \left(k_{u} \psi_{u}(\vect{x}) - k_{v} \psi_{v}(\vr+\vect{x})\right)} \right. \nonumber \\
& \left. \left\lbrace \delta_{g}^{r}(\vr) + \partial_{u} \psi_{u}(\vr) \right\rbrace \left\lbrace \delta_{g}^{r}(\vr+\vect{x}) + \partial_{v} \psi_{v}(\vr+\vect{x}) \right\rbrace \right\rangle.
\label{eq:los_cross_power_spectrum}
\end{align}
The real-space galaxy density field at position $\vr$ is denoted as $\delta_{g}^{r}(\vr)$, and we have used 
$k_{u} = \vect{k} \cdot u = k \cos \theta_{u}$, where $\cos \theta_{u}$ is the cosine of the angle 
between the wave vector $\vect{k}$ and line of sight $u$ (with similar definitions for $k_{v}$ and
$\cos \theta_{v}$). When $\vert k \psi_{u}\vert, \vert k \psi_{v}\vert \ll 1$, and using that, at 
linear order, $i k_{u} \psi_{u} = \cos^{2}\theta_{u} f \delta_{m}$ (with $\delta_{m}$ the 
real-space matter density contrast), we find the equivalent of the Kaiser formula \citep{Kaiser1987}
for two lines of sight,
\begin{align}
& \left\langle \delta_{g}^{u}(\vect{k})\delta_{g}^{v}(\vect{k}^{\prime}) \right\rangle \nonumber \\
& = (2\pi)^3 \delta^D(\vect{k}+\vect{k}^{\prime}) b^{2} \left(1 + \beta \cos^{2} \theta_{u}\right) \left(1 + \beta \cos^{2} \theta_{v}\right) P_{m}^{\mathrm{lin}}(k).
\label{eq:cross_density_contrast}
\end{align}
Here, we have assumed a linear galaxy bias, $b$, (i.e. $\delta_{g}^{r} = b \delta_{m}$), and define 
$\beta = f/b$. $P_{m}^{\mathrm{lin}}(k)$ is the linear matter power spectrum.
}

For the case in which the two lines of sight are perpendicular (which we emphasise by setting $u=x$ and $v=y$),
we can put these ingredient together, using $\cos{\theta_{y}} = \sin{\theta_{x}} \cos{\phi_{x}}$
and $d\Omega_{x} = d\phi_{x}d\theta_{x}\sin{\theta_{x}}$. Finally, $\rho_{\ell\ell^{\prime}ij}^{xy}$
can be written as
\begin{equation}
\label{eq:cross_correlation_orthogonal}
\rho_{\ell\ell^{\prime}ij}^{xy}(\beta) = \delta_{ij}^{K}\frac{\kappa_{\ell \ell^{\prime}}^{xy}(\beta)}{\kappa_{\ell \ell^{\prime}}(\beta)},
\end{equation}
where
\begin{align}
\kappa_{\ell \ell^{\prime}}^{xy}(\beta) = &\int d\phi_{x}d\theta_{x}\sin{\theta_{x}} \left(1 + \beta \cos^{2}{\theta_{x}}\right)^{2} \nonumber \\ 
& \left(1 + \beta \sin^{2}{\theta_{x}} \cos^{2}{\phi_{x}} \right)^{2} \Leg{\ell}(\cos{\theta_{x}})\Leg{\ell^{\prime}}(\sin{\theta_{x}}\cos{\phi_{x}}), \label{eq:cross_correlation_orthogonal_los} \\
\kappa_{\ell \ell^{\prime}}(\beta) = & 2\pi \int d\theta_{x}\sin{\theta_{x}} \left(1 + \beta \cos^{2}{\theta_{x}}\right)^{4} \Leg{\ell}(\cos{\theta_{x}})\Leg{\ell^{\prime}}(\cos{\theta_{x}}).
\label{eq:auto_correlation_orthogonal_los}
\end{align}

\begin{figure}
\centering
\includegraphics[width=\columnwidth]{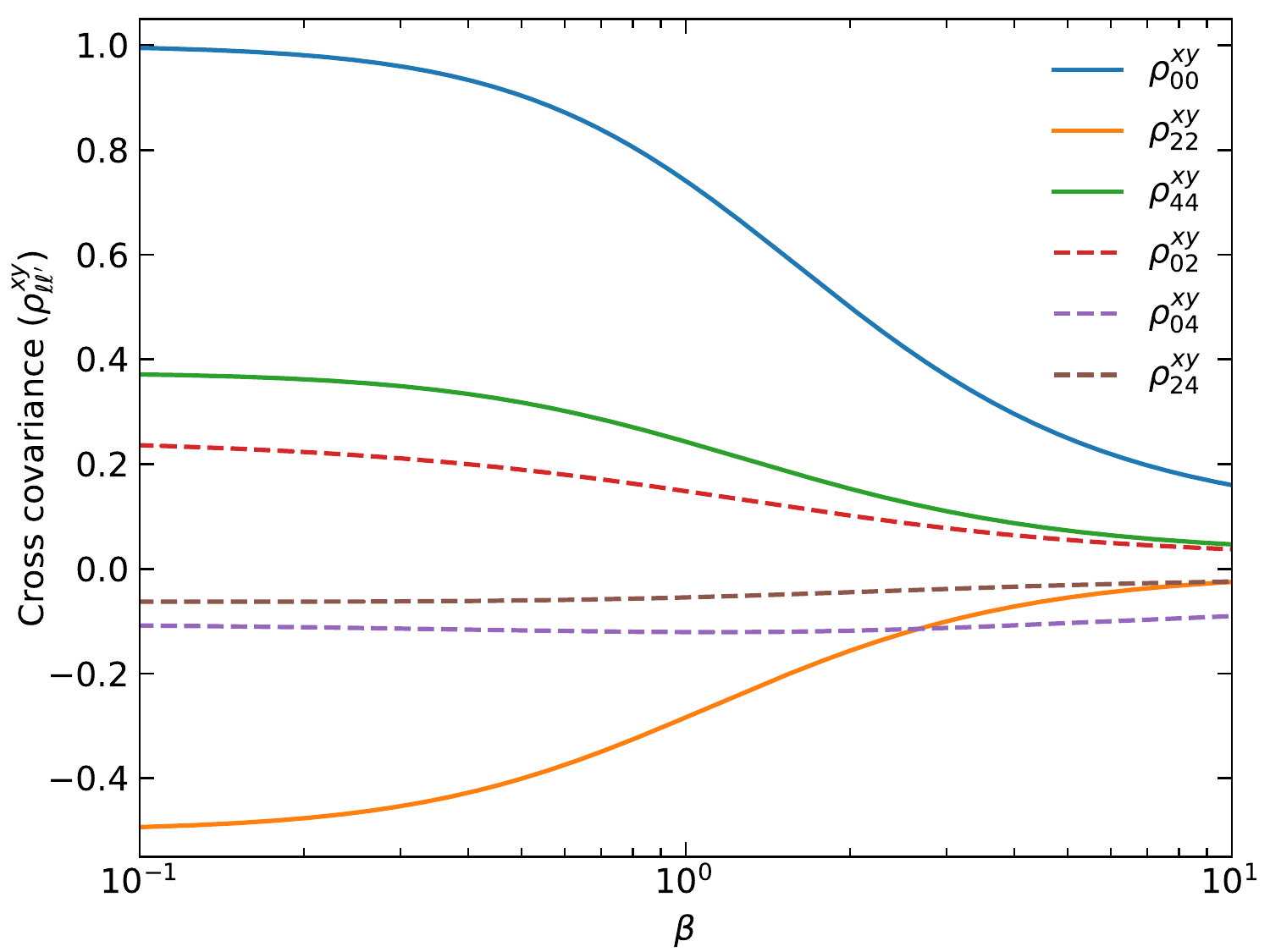}
\caption{\highlight{Cross-correlation} $\rho^{xy}_{\ell\ell^{\prime}}$, calculated analytically for the power spectrum 
multipoles. Cross-correlations are shown by the solid curves for the monopole 
($\rho^{xy}_{00}$, blue), quadrupole ($\rho^{xy}_{22}$, yellow) and hexadecapole 
($\rho^{xy}_{44}$, green), with two orthogonal lines of sight,
assuming no shot noise. The cross terms are indicated by the dashed curves, with 
$\rho^{xy}_{02}$ (red), $\rho^{xy}_{04}$ (purple) and $\rho^{xy}_{24}$ (brown).}
\label{fig:correlation_los}
\end{figure}

Here, where we have neglected shot noise, the \highlight{cross-correlation} only depends on $\beta$, 
and is independent of $k$.
$\rho_{\ell\ell^{\prime}}^{xy}$ is shown in Fig.~\ref{fig:correlation_los}.
For the monopole ($\rho_{00}^{xy}$), the cross-correlation reaches $1$ if there is no RSD 
($\beta=0$), which is as expected, and decreases as $\beta$ is increased. 
For the quadrupole ($\rho_{22}^{xy}$), the cross-correlation is negative, which indicates
that there is an anti-correlation between the quadrupole measurements for two lines of sight.

The limits as $\beta \rightarrow 0$ and $\beta \rightarrow \infty$ for the different multipoles are
\begin{align}
\lim_{\beta \rightarrow 0} \rho_{00ii}^{xy}(\beta) = 1 \qquad & \lim_{\beta \rightarrow +\infty}\rho_{00ii}^{xy}(\beta) = \frac{3}{35} \nonumber \\
\lim_{\beta \rightarrow 0} \rho_{22ii}^{xy}(\beta) = -\frac{1}{2} \qquad & \lim_{\beta \rightarrow +\infty} \rho_{22ii}^{xy}(\beta) = -\frac{33}{5810} \\
\lim_{\beta \rightarrow 0} \rho_{44ii}^{xy}(\beta) = \frac{3}{8} \qquad & \lim_{\beta \rightarrow +\infty} \rho_{44ii}^{xy}(\beta) = \frac{5239}{199080} \nonumber.
\end{align}

These cross-correlations are related to the scatter between the power spectrum 
(and correlation function) measurements on large scales (as shown in Fig.~\ref{fig:pk_xi} for 
the OuterRim haloes). Very little scatter is seen in the monopole measurements with 
orthogonal lines of sight since they are highly correlated, while the variations are much
larger for the quadrupole, where the measurements are anti-correlated.

In Section~\ref{sec:velocities}, we showed that the relative amplitude of the
quadrupole measurements were related to the rms of the velocity distributions. 
This is not the full picture, since they are also affected by the cross-correlations.
In the limit of no RSD, the amplitude of the quadrupole goes to zero, but the
anti-correlation is at a maximum, and variations are seen due to pairs being assigned
to different $\mu$ bins when being observed from different directions. When RSD
is included, the variations in the quadrupole are due to the velocity distributions.
When the velocity is increased, the overall amplitude of the quadrupole increases,
but the relative difference between the measurements decreases, due to a 
reduction in the anti-correlation. Despite this, it is still the direction where the
rms of the velocity distribution is largest that will have the strongest quadrupole.

\subsection{General line of sight and shot noise}
\label{sec:cross_covariance_shot_noise}

The \highlight{cross-correlation} derived in the previous section can be generalised for any
two lines of sight, which are not necessarily orthogonal.
Taking the line of sight $u$ as being directed along the reference $x$-axis ($\vect{\hat{u}}=(1,0,0)$), we can write the
unit wavevector as $\hat{k} = \left(\cos\theta_{u}, \sin\theta_{u} \cos\phi_{u}, \highlight{\sin\theta_{u} \sin\phi_{u} }\right)$. 
As before, $\theta_u$ is the angle between $\vect{\hat{k}}$ and $\vect{\hat{u}}$, and $\phi_u$ is the azimuthal angle around $u$. 
In the same coordinate system we can also write the second line of sight, $v$, as 
$\vect{\hat{v}} = \left(\cos\theta_{uv},\sin\theta_{uv} \cos\phi_{uv}, \highlight{\sin\theta_{uv} \sin\phi_{uv}} \right)$, where 
$\theta_{uv}$ is the angle between $v$ and $u$, and $\phi_{uv}$ is the azimuthal angle of $v$ around $u$.
The cosines of the angles, $\cos \theta_{u} = \hat{k} \cdot \vect{\hat{u}}$ and $\cos \theta_{v} = \hat{k} \cdot \vect{\hat{v}}$, 
are therefore related through
\begin{align}
\cos\theta_{v} &= \cos\theta_{u} \cos\theta_{uv} + \sin\theta_{u} \cos\phi_{u} \sin\theta_{uv} \cos\phi_{uv} \nonumber \\
&+ \highlight{\sin\theta_{u} \sin\phi_{u} \sin\theta_{uv} \sin\phi_{uv}} \nonumber \\
&= \cos\theta_{u} \cos\theta_{uv} + \sin\theta_{u} \sin\theta_{uv} \cos\phi_{v},
\end{align}
where, in the last equality, we have used that $\phi_{v} = \phi_{uv}-\phi_{u}$.

\highlight{
So far, we have neglected the shot noise coming for the discreteness of the galaxy density field. 
The shot noise contribution to the $\aver{e^{i \left(k_{u} \psi_{u}(\vr) - k_{v} \psi_{v}(\vr+\vect{x})\right)} \delta_{g}^{r}(\vr) \delta_{g}^{r}(\vr+\vect{x})}$ 
term of Eq.~\ref{eq:los_cross_power_spectrum} reads $\aver{e^{i \left(k_{u} \psi_{u}(\vr) - k_{v} \psi_{v}(\vr+\vect{x})\right)} \left(1+\delta_{g}^{r}(\vr)\right)} \delta_{D}^{(3)}(\vect{x})/\bar{n}_{g}$. 
Performing the integral over $\vect{x}$ gives
\begin{equation}
P^{\mathrm{noise}}(\vect{k}) = \left\langle \frac{1}{\bar{n}_{g}}e^{i \left( k_{u} \psi_{u} - k_{v} \psi_{v} \right)}\left(1+\delta_{g}^{r}\right) \right\rangle,
\label{eq:shot_noise}
\end{equation}
where $\delta_{g}^{r}$, $\psi_{u}$ and $\psi_{v}$ are taken at the same configuration-space position. 
The exponential $e^{i \left( k_{u} \psi_{u} - k_{v} \psi_{v} \right)}$ can be expanded into a series, and 
after applying Wick's theorem, yields products involving $\big\langle\psi_{u} \delta_{g}^{r} \big\rangle$ 
or $\big\langle \psi_{v} \delta_{g}^{r}\big\rangle$, which are zero at linear order, as we will see in 
the following. Hence $\aver{e^{i \left( k_{u} \psi_{u} - k_{v} \psi_{v} \right)}\delta_{g}^{r}} = 0$, 
i.e. the $\delta_{g}^{r}$ term can be dropped in Eq.~\ref{eq:shot_noise}. Next, the phase shift term 
$\aver{ e^{i \left( k_{u} \psi_{u} - k_{v} \psi_{v} \right)} }$ is the characteristic function of 
the Gaussian random variable, $k_{u} \psi_{u} - k_{v} \psi_{v}$, which has a mean of zero, 
and variance $\sigma_{\psi,uv}^{2}$.
}

\highlight{
Therefore the shot noise contribution reads
\begin{equation}
P^{\mathrm{noise}}(\vect{k}) = \frac{1}{\bar{n}_{g}}e^{-\frac{1}{2}\sigma_{\psi,uv}^{2}(k)},
\end{equation}
where $\sigma_{\psi,uv}^{2}(k)$ is given by
\begin{align}
\sigma_{\psi,uv}^{2}(k) &= \big\langle\left( k_{u} \psi_{u} - k_{v} \psi_{v} \right)^{2}\big\rangle \\
&= k_{u}^{2} \big\langle \psi_{u}^{2}\big\rangle + k_{v}^{2} \big\langle \psi_{v}^{2} \big\rangle - 2 k_{u}k_{v} \big\langle \psi_{u}\psi_{v} \big\rangle.
\end{align}
}
From linear perturbation theory, the displacement $\psi_u$ due to redshift space distortions is given in Fourier space by $\psi_u(\vect{k}) = i f (k_u / k^2) \delta(\vect{k})$. We therefore have, taking the Fourier transform,
\begin{align} \label{eq:psi_cross_correlation}
\big\langle \psi_{u}\psi_{v} \big\rangle &= \frac{f^{2}}{\left(2\pi\right)^{3}} \int d^{3} k \frac{k_{u}k_{v}}{k^{4}} P_m^\mathrm{lin}(k) \nonumber \\
&= \frac{f^{2}}{\left(2\pi\right)^{3}} \int d k P_m^\mathrm{lin}(k) \int_{0}^{2\pi} d\phi_{v} \nonumber \\
& \int_{0}^{\pi} \sin\theta_{u} d\theta_{u} \cos\theta_{u} \nonumber \\
& \left( \cos\theta_{u} \cos\theta_{uv} + \sin\theta_{u} \sin\theta_{uv} \cos\phi_{v} \right) \nonumber \\
&= f^{2} \sigma_{d}^{2} \cos\theta_{uv},
\end{align}
where the one-dimensional rms of the displacement field is
\begin{equation}
\sigma_{d}^{2} = \frac{1}{6 \pi^2} \int d k P_m^\mathrm{lin}(k).
\end{equation}
Since $\cos\theta_{uu} = \cos\theta_{vv} = 1$, we also have
$\left\langle\psi_{u}^{2}\right\rangle = \left\langle\psi_{v}^{2}\right\rangle = f^{2} \sigma_{d}^{2}$, and therefore, $\sigma_{\psi,uv}^{2}(k) = \left(k_{u}^{2} + k_{v}^{2} - 2 k_{u}k_{v}  \cos\theta_{uv}\right) f^{2} \sigma_{d}^{2}$.
In the case that the two lines of sight are orthogonal ($\cos\theta_{uv} = 0$), $\psi_{u}$ and $\psi_{v}$ do 
not correlate. 
\highlight{Note also that $\big\langle \psi_{u}\delta_{g}^{r} \big\rangle$ yields an integral over $d\theta_{u} \cos\theta_{u}$ that is zero, which justifies dropping $\delta_{g}^{r}$ in Eq.~\ref{eq:shot_noise}, as discussed previously. }
The shot noise contribution is largest ($1/\bar{n}_{g}$) when $f = 0$, as expected,
and vanishes as $f \rightarrow \infty$.

Finally, putting everything together, and using $d\Omega_{u} = d\phi_{v}d\theta_{u}\sin{\theta_{u}}$, 
the \highlight{cross-correlation} $\rho_{\ell\ell^{\prime}ij}^{uv}(\beta, b, \bar{n}_{g})$ can be written as 
\begin{equation}
\label{eq:cross_correlation_shot}
\rho_{\ell\ell^{\prime}ij}^{uv}(\beta, b, \bar{n}_{g}) = \delta_{ij}^{K} \frac{\kappa_{\ell \ell^{\prime}}^{uv}(k_{i},\beta,b,\bar{n}_{g})}{\kappa_{\ell \ell^{\prime}}(k_{i},\beta,b,\bar{n}_{g})},
\end{equation}
with, at leading order in $(\bar{n}_{g}P_m^\mathrm{lin}(k))^{-1}$ \citep[e.g.][]{Meiksin1999,Howlett2017}, 
\begin{align}
\kappa_{\ell \ell^{\prime}}^{uv}(k,\beta,b,\bar{n}_{g}) &= \int_{0}^{2\pi} d\phi_{v} \int_{0}^{\pi} d\theta_{u}\sin{\theta_{u}} \nonumber \\
& \left[b^{2}\left(1 + \beta \cos^{2}{\theta_{u}}\right) \left(1 + \beta \cos^{2}{\theta_{v}}\right) P_m^\mathrm{lin}(k) \right. \nonumber \\
& \left. + \frac{1}{\bar{n}_{g}} e^{-k^{2}\left(\cos^{2}{\theta_{u}} + \cos^{2}{\theta_{v}} - 2 \cos{\theta_{u}}\cos{\theta_{v}}  \cos \theta_{uv} \right) f^{2} \sigma_{d}^{2}/2} \right]^{2} \nonumber \\
 & \Leg{\ell}(\cos{\theta_{u}})\Leg{\ell^{\prime}}(\cos\theta_{v}), \label{eq:cross_correlation_shot_los} \\
\kappa_{\ell \ell^{\prime}}(k,\beta,b,\bar{n}_{g}) &= 2\pi \int_{0}^{\pi} d\theta_{u}\sin{\theta_{u}} \left[ b^{2} \left(1 + \beta \cos^{2}{\theta_{u}}\right)^{2} P_m^\mathrm{lin}(k) + \frac{1}{\bar{n}_{g}} \right]^{2} \nonumber \\
& \Leg{\ell}(\cos{\theta_{u}})\Leg{\ell^{\prime}}(\cos{\theta_{u}}).
\label{eq:auto_correlation_shot_los}
\end{align}
In the limit that there is no shot noise, $1/\bar{n}_{g} \rightarrow 0$, and the $P_m^\mathrm{lin}(k)$
in Eq.~\ref{eq:cross_correlation_shot_los} cancels with the $P_m^\mathrm{lin}(k)$ in 
Eq.~\ref{eq:auto_correlation_shot_los}, removing any $k$-dependence on the \highlight{cross-correlation}.
If the lines of sight $u$ and $v$ are orthogonal, then we arrive at the
same expression as Eq.~\ref{eq:cross_correlation_orthogonal_los}~\&~\ref{eq:auto_correlation_orthogonal_los}.

\subsection{Gain in uncertainty averaging over multiple lines of sight}
\label{sec:gain_uncertainty_theory}

The reduction in the errors in the power spectrum measurements when averaging together 
multiple lines of sight is related to the \highlight{cross-correlation} derived in the previous subsections. 
For example, as seen for the OuterRim simulation in 
Section~\ref{sec:clustering_statistics_los}, the monopole measurements are highly correlated,
resulting in only a small gain. However, quadrupole measurements are anti-correlated
resulting in very large gains in the uncertainties.
Here, we show the relationship between the \highlight{cross-correlation} and the improvement
in the uncertainties of the power spectrum measurements.

\subsubsection{Two lines of sight}

If we consider the case of 2 orthogonal lines of sight, the mean power spectrum 
measurement is simply
\begin{equation}
\hat{P}^\mathrm{2-los} = \frac{\hat{P}^x+\hat{P}^y}{2}.
\end{equation}
Its covariance matrix, $C$, (\highlight{where we drop the indices for conciseness, and without loss of
generality take the power spectrum mean to be zero}) is
\begin{align}
\left\langle \Big(\hat{P}^\mathrm{2-los}\Big)^2\right\rangle &= \left\langle \Big( \frac{\hat{P}^x+\hat{P}^y}{2} \Big)^2 \right\rangle \\
C \rho^\mathrm{2-los} &= \frac{1}{4}\left[ \langle \hat{P}^x \hat{P}^x \rangle + \langle \hat{P}^y \hat{P}^y \rangle + 2 \langle \hat{P}^x \hat{P}^y \rangle \right]\\
&= \frac{1}{2} \left[ \langle \hat{P}^x \hat{P}^x \rangle + \langle \hat{P}^x \hat{P}^y \rangle \right]\\
&= \frac{C}{2} (1+\rho^{xy}),
\end{align}
since $\langle \hat{P}^x \hat{P}^x\rangle = \langle \hat{P}^y \hat{P}^y\rangle = C$, and 
$\langle \hat{P}^x \hat{P}^y\rangle = C\rho^{xy}$, where $\rho^{xy}$ is the \highlight{cross-correlation}.
The reduction in the variance of the average measurement is therefore
\begin{equation}
\rho^\mathrm{2-los} = \frac{1+\rho^{xy}}{2},
\end{equation}
and the reduction in the uncertainty is $\sqrt{\rho^\mathrm{2-los}}$.

\subsubsection{Three lines of sight}

The above can be extended to three orthogonal lines of sight by writing the
average measurement as 
\begin{equation}
\hat{P}^\mathrm{3-los} = \frac{\hat{P}^x+\hat{P}^y+\hat{P}^z}{3}.
\end{equation}
Following the same logic, it can be shown that the gain in the variance is
\begin{equation}
\label{eq:variance_reduction_3los}
\rho^\mathrm{3-los} = \frac{1+2\rho^{xy}}{3}.
\end{equation}
For the case of no shot noise, the limits are 
\begin{align}
\lim_{\beta \rightarrow 0} \rho_{00ii}^{\mathrm{3-los}}(\beta) = 1 \qquad & \lim_{\beta \rightarrow +\infty}\rho_{00ii}^{\mathrm{3-los}}(\beta) = \frac{41}{105} \nonumber \\
\lim_{\beta \rightarrow 0} \rho_{22ii}^{\mathrm{3-los}}(\beta) = 0 \qquad & \lim_{\beta \rightarrow +\infty} \rho_{22ii}^{\mathrm{3-los}}(\beta) = \frac{2872}{8715} \nonumber \\
\lim_{\beta \rightarrow 0} \rho_{44ii}^{\mathrm{3-los}}(\beta) = \frac{7}{12} \qquad & \lim_{\beta \rightarrow +\infty} \rho_{44ii}^{\mathrm{3-los}}(\beta) = \frac{104779}{298620} \nonumber.
\end{align}

In the limit as $\beta \rightarrow 0$, the relative variance of the monopole is 1,
as expected, since the three monopole measurements in real space are identical. 
For the quadrupole, the relative variance is reduced to 0.

The predicted gain in the uncertainties ($\sqrt{\rho^\mathrm{3-los}(\beta)}$) for the 
multipole measurements is shown by the solid curves in Fig.~\ref{fig:error_gain},
for the case of 3 orthogonal lines of sight, with no shot noise.
We also measure the gain in uncertainties from a set of 100 Gaussian
random fields, shown by the points, where redshift-space distortions are added using the Kaiser
formula with different values of $\beta$. The measured gain in the uncertainties
from the Gaussian random fields is in good agreement with the prediction.
The red vertical dashed line indicates the value of $\beta$ for the 
OuterRim halo catalogue, and the gain is consistent with the gain
seen on large scales in Fig.~\ref{fig:pk_xi_err}.

\subsubsection{All lines of sight}

One can go further and average over all possible lines of sight, $\omega$, of the solid angle $\Omega$, 
\begin{equation}
\hat{P}^{\mathrm{all-los}} = \frac{1}{4\pi} \int d\Omega \hat{P}^{\omega}.
\end{equation}
As before, we have
\begin{equation}
\aver{\left( \hat{P}^{\mathrm{all-los}} \right)^{2}} = \frac{1}{\left(4\pi\right)^{2}} \int d\Omega \int d\Omega^{\prime} \aver{\hat{P}^{\omega}\hat{P}^{\omega^{\prime}}}.
\end{equation}
Following from statistical isotropy, a reference line of sight $u$ can be chosen, such that
\begin{equation}
\aver{\left( \hat{P}^{\mathrm{all-los}} \right)^{2}} = \frac{1}{4\pi} \int d\Omega \aver{\hat{P}^{u} \hat{P}^{\omega}},
\end{equation}
and therefore, when averaging over all lines of sight, the gain in the variance is
\begin{equation}
\rho^\mathrm{all-los} = \frac{1}{2} \int^\pi_0 d\theta_{uv} \sin \theta_{uv} \rho^{uv}.
\end{equation}

Already, with $3$ orthogonal lines of sight, the relative variance of the quadrupole is pinned down to $0$ in the limit $\beta \rightarrow 0$. When averaging over all lines of sight, the limits are
\begin{align}
\lim_{\beta \rightarrow 0} \rho_{00ii}^{\mathrm{all-los}}(\beta) = 1 \qquad & \lim_{\beta \rightarrow +\infty}\rho_{00ii}^{\mathrm{all-los}}(\beta) = \frac{9}{25} \nonumber \\
\lim_{\beta \rightarrow 0} \rho_{22ii}^{\mathrm{all-los}}(\beta) = 0 \qquad & \lim_{\beta \rightarrow +\infty} \rho_{22ii}^{\mathrm{all-los}}(\beta) = \frac{20592}{101675} \nonumber \\
\lim_{\beta \rightarrow 0} \rho_{44ii}^{\mathrm{all-los}}(\beta) = 0 \qquad & \lim_{\beta \rightarrow +\infty} \rho_{44ii}^{\mathrm{all-los}}(\beta) = \frac{155584}{7838775} \nonumber. \\
\end{align}
This also pins down the relative variance of the hexadecapole to $0$ 
as $\beta \rightarrow 0$, while there is also a reduction in the limits as 
$\beta \rightarrow \infty$ for all multipoles.

\begin{figure}
\centering
\includegraphics[width=\columnwidth]{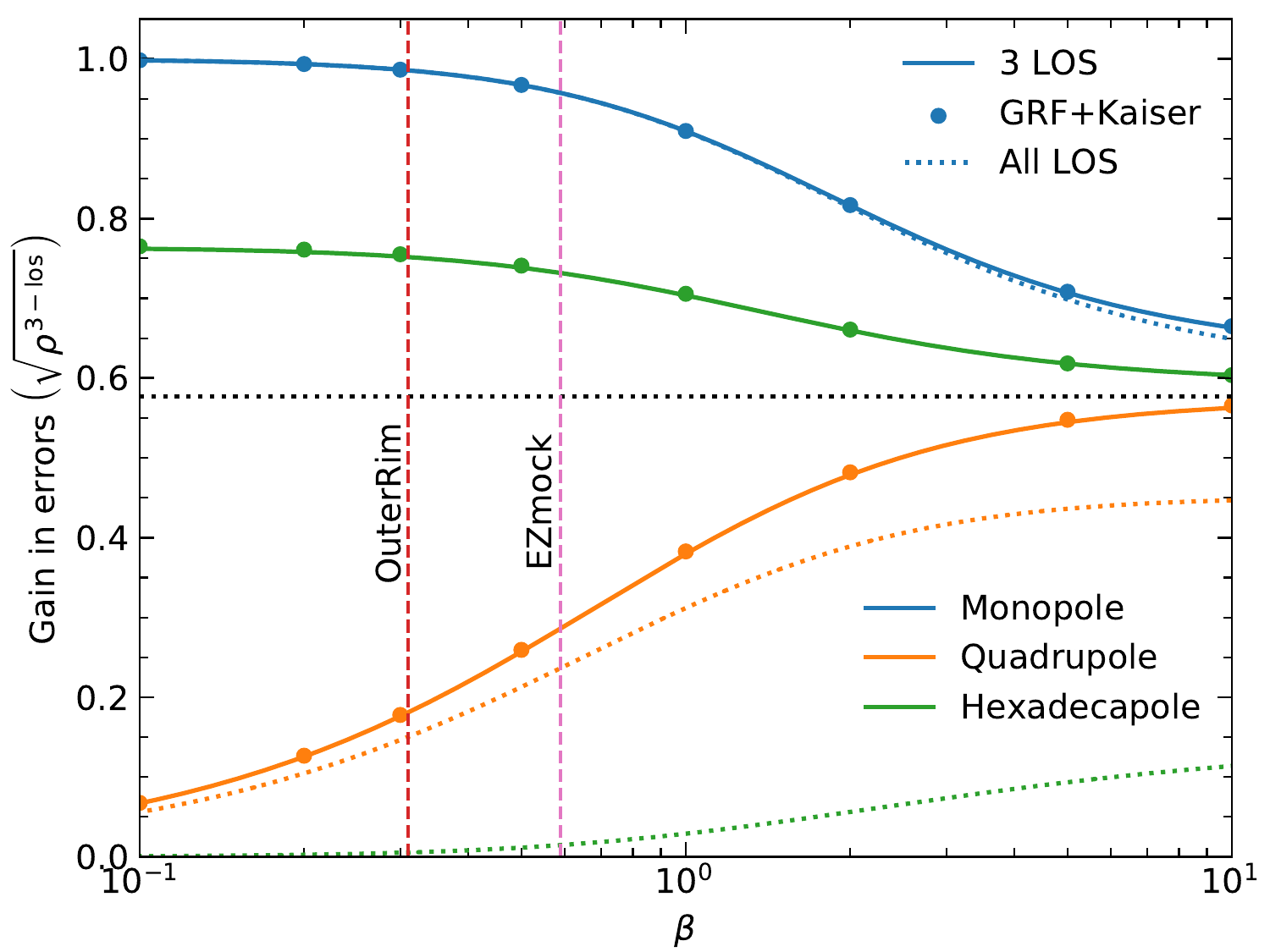}
\caption{Gain in the uncertainties, $\sqrt{\rho^\mathrm{3-los}}$, for the power spectrum
multipoles when averaging measurements over multiple lines of sight.
The solid curves show the prediction for 3 orthogonal lines of sight, 
of the monopole (blue), quadrupole (yellow), and hexadecapole (green), assuming zero shot noise.
Points are measured from 100 Gaussian random fields, with RSD applied.
Dotted curves indicate the case where the multipoles are averaged over all lines of sight.
The red and pink vertical dashed lines indicate the values of $\beta$ for the
OuterRim halo catalogue, and for the EZmock boxes. The horizontal
black dashed line indicates a factor of $1/\sqrt{3}$.}
\label{fig:error_gain}
\end{figure}

The gain in the uncertainties when averaging all lines of sight is shown by the
dotted curves in Fig.~\ref{fig:error_gain}. Compared to the case of 3 orthogonal lines,
there is very little change in the monopole, with some improvement in the quadrupole.
For the hexadecapole, the improvement is much more striking, where the relative gain
in the uncertainties is pulled very close to zero even at large values of $\beta$.
While averaging over all lines of sight greatly improves the hexadecapole
measurement, this comes at the cost of inflating the cross-terms in the
covariance matrix. The monopole and quadrupole of the power spectrum averaged over all lines of sight (and also the monopole and hexadecapole) are fully correlated, such that the covariance matrix is singular.

\begin{figure}
\centering
\includegraphics[width=\columnwidth]{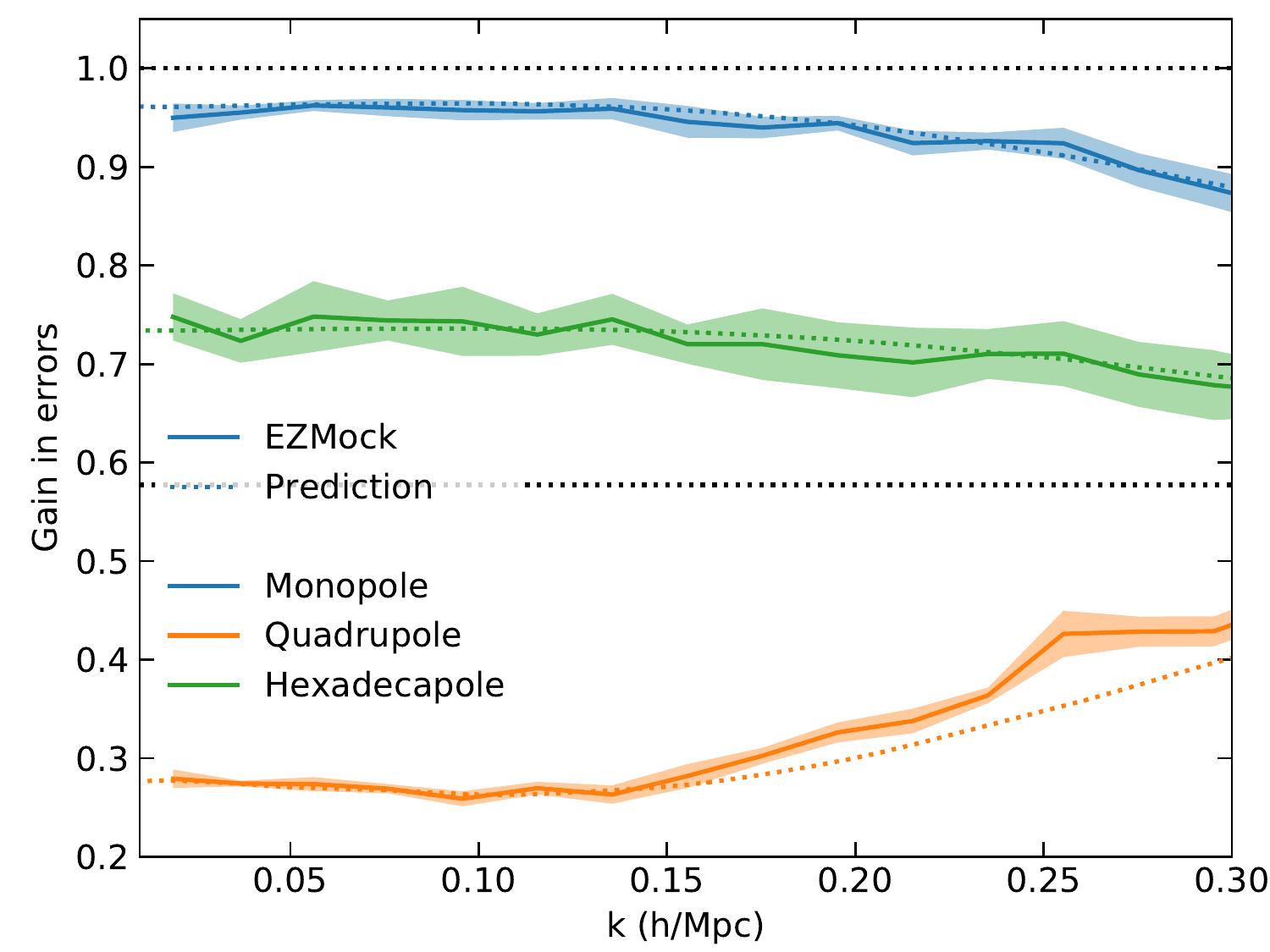}
\caption{Gain in the uncertainties of the power spectrum multipoles from the
set of 300 EZmocks. The gain measured from the mocks is shown by the solid
curves for the monopole (blue), quadrupole (yellow) and hexadecapole (green).
The error on the error is estimated by splitting the 300 mocks into 6 sets of 
50 mocks. The prediction, which includes shot noise, is shown by the dotted curves.}
\label{fig:error_gain_ezmock}
\end{figure}

\subsection{Discussion}
\label{sec:discussion}

\highlight{
We have shown that in linear theory, averaging together power spectra measurements over multiple
lines of sight can result in large gains in the multipole uncertainties. Here, we provide a more
intuitive explanation for where this additional information comes from.
}

\highlight{
All the information for a potential flow is contained within the velocity divergence, $\theta$.
However, this information cannot be extracted from the velocity field along a single line of sight.
Since the line-of-sight velocity $v_s(k,\mu) \propto \mu \theta(k)/k$ at large scales, the transverse modes are
suppressed by a factor of $\mu$. It is therefore only possible to obtain precise measurements 
of $\theta$ along the line of sight direction. In addition, for RSD, we measure 
$\delta(\vect{k}) + f \mu^2 \theta(\vect{k})$, where $\theta$ is suppressed by an additional factor of $\mu$.
Adding a second line of sight restores sensitivity to other $\vect{k}$ modes of $\theta$, 
and the most information is added when the lines of sight are orthogonal. 
}

\highlight{From Eq.~\ref{eq:psi_cross_correlation}, the cross-correlation between the displacements
for two lines of sight, $\psi_{u}$ and $\psi_{v}$, is not $1$ and varies with the angle between them, 
$\cos\theta_{uv}$. This means that redshift-space density contrasts along the two different lines of sight
(such that $\cos\theta_{uv} \neq \pm 1$) are not fully correlated, explaining why averaging the lines
of sight produces a small reduction in the uncertainties of the monopole in Section~\ref{sec:gain_uncertainties_measured}.}

\highlight{Averaging over multiple lines of sight allows much more accurate measurements to be
taken of the velocity dispersion, since the multiple lines of sight enable other $\vect{k}$ modes to
be probed. As discussed in Section~\ref{sec:velocities}, the amplitude of the quadrupole is
related to the velocity dispersion, so greatly improving the accuracy of the velocity dispersion
estimates results in much more precise measurements of the quadrupole.}


\section{Fisher analysis}

\label{sec:fisher_analysis}

In this section, we propagate the cross-covariance between power spectrum measurements down to the 
measurements of the cosmological parameters.

\subsection{Theory}

Cosmological parameters are estimated through a $\chi^{2}$ minimisation,
\begin{equation}
\chi^{2}(\vect{p}) = \left[\vect{P}-\vect{M}(\vect{p})\right]^{T} \vect{C}^{-1} \left[\vect{P}-\vect{M}(\vect{p})\right],
\end{equation}
where $\vect{P}$ is the measured power spectrum, and $\vect{M}(\vect{p})$ is the model, 
which depends on parameters $\vect{p}$, and $\vect{C}$ is the covariance matrix. 
Taking the derivative of $\chi^{2}(\vect{p})$ with
respect to $\vect{p}$ yields
\begin{equation}
\vect{0} =
\left(\frac{\partial \vect{M}(\vect{p})}{\partial \vect{p}}\right)^{T} \vect{C}^{-1} \left[\vect{P}-\vect{M}(\hat{\vect{p}})\right].
\label{eq:chi2_differentiated}
\end{equation}
At first order around the true parameter value $\vect{p}_0$ we can write $\vect{M}(\hat{\vect{p}}) = \vect{M}(\vect{p}_0) + \frac{\partial \vect{M}(\vect{p})}{\partial \vect{p}} \cdot \left(\hat{\vect{p}}-\vect{p}_0\right)$, with all derivatives with respect to $\vect{p}$ taken at $\vect{p}_0$. Without loss of generality, we set $\vect{p}_0 = \vect{0}$ and $\vect{M}(\vect{p}_0) = \vect{0}$. Eq.~\ref{eq:chi2_differentiated} can therefore be rearranged to give
\begin{equation}
\left(\frac{\partial \vect{M}(\vect{p})}{\partial \vect{p}}\right)^{T} \vect{C}^{-1} \left(\frac{\partial \vect{M}(\vect{p})}{\partial \vect{p}}\right) \hat{\vect{p}} = \left(\frac{\partial \vect{M}(\vect{p})}{\partial \vect{p}}\right)^{T} \vect{C}^{-1} \vect{P}.
\end{equation}
Writing the Fisher information as
\begin{equation}
\vect{F} = \left(\frac{\partial \vect{M}(\vect{p})}{\partial \vect{p}}\right)^{T} \vect{C}^{-1} \left(\frac{\partial \vect{M}(\vect{p})}{\partial \vect{p}}\right),
\end{equation}
we obtain
\begin{equation}
\hat{\vect{p}} = \vect{F}^{-1} \left(\frac{\partial \vect{M}(\vect{p})}{\partial \vect{p}}\right)^{T} \vect{C}^{-1} \vect{P}.
\end{equation}
The covariance of the measured parameters, when the power spectra measurements are calculated 
with respect to the same line of sight, is simply
\begin{align}
\aver{\hat{\vect{p}} \hat{\vect{p}}^{T}} &= \vect{F}^{-1} \left(\frac{\partial \vect{M}(\vect{p})}{\partial \vect{p}}\right)^{T} \vect{C}^{-1} \aver{\vect{P}\vect{P}^{T}} \vect{C}^{-1} \left(\frac{\partial \vect{M}(\vect{p})}{\partial \vect{p}}\right) \vect{F}^{-1} \nonumber \\
&= \vect{F}^{-1}.
\end{align}
For two different lines of sight $u$ and $v$, the covariance is
\begin{align}
\aver{\hat{\vect{p}}^{u} \hat{\vect{p}}^{v,T}} &= \vect{F}^{-1} \left(\frac{\partial \vect{M}(\vect{p})}{\partial \vect{p}}\right)^{T} \vect{C}^{-1} \aver{\vect{P}^{u}\vect{P}^{v,T}} \vect{C}^{-1} \left(\frac{\partial \vect{M}(\vect{p})}{\partial \vect{p}}\right) \vect{F}^{-1} \nonumber \\
&= \vect{F}^{-1} \left(\frac{\partial \vect{M}(\vect{p})}{\partial \vect{p}}\right)^{T} \vect{C}^{-1} \vect{C}^{uv} \vect{C}^{-1} \left(\frac{\partial \vect{M}(\vect{p})}{\partial \vect{p}}\right) \vect{F}^{-1},
\label{eq:fisher_forecast_cross}
\end{align}
where $\vect{C}^{uv}$ is the cross-covariance between the two different lines of sight $u$ and $v$ for 
measurements of the power spectrum, which is estimated in Section~\ref{sec:cross_covariance_theory}. 
The same calculations
derived in Section~\ref{sec:gain_uncertainty_theory} can be used to calculate the
gain in variance on the cosmological parameters, when averaging over multiple lines of sight.

\subsection{Measurements from EZmocks}

We test the Fisher forecasts by comparing the gain in uncertainties of the cosmological parameters against measurements from the eBOSS ELG EZmock boxes described in Section~\ref{sec:ezmocks}.

The power spectrum in the periodic box is measured for the 300 EZmocks, with RSD applied along the $x$, $y$ and $z$-directions.
We show the gain in the uncertainties in the power spectrum measurements, when averaging together the three lines of sight, in Fig.~\ref{fig:error_gain_ezmock}, where the uncertainty is measured from the error on the mean. 
To estimate the error on the error, the 300 mocks are split into
6 sets of 50, and the standard deviation is calculated between the errors evaluated for each set. The dotted lines show the theoretical prediction (from Eqs.~\ref{eq:cross_correlation_shot}-\ref{eq:auto_correlation_shot_los}), where $\beta$ and $b$ are the mean values obtained by fitting the measured power spectrum monopoles, quadrupoles and hexadecapoles with a linear Kaiser model damped by a Lorentzian Finger-of-God term over the $k$-range $0.03<k<0.12~\invhMpc$ (see below). $f \sigma_d$ is taken to be the rms of the RSD displacement measured in the mocks.
Since the mocks are affected by shot noise, the gain in uncertainties is a function of $k$, with the biggest gains on large, linear scales (small $k$). This gain is consistent with the prediction, given the value of $\beta$ of
the EZmocks, as indicated by the pink vertical dashed line in Fig.~\ref{fig:error_gain}. The relative uncertainty, as a function of $k$, is mostly in good agreement with the
prediction. On small scales, there is some small disagreement
in the quadrupole, where the measured gain is smaller than predicted. However, since our prediction assumes linear theory, we expect it will break down on small scales, where non-linearities become important.

A model of the redshift-space power spectrum is fit to the measured power spectra of each mock, in order to obtain the best fit values of the cosmological parameters. 
\highlight{The fitting procedure is performed individually on each line of sight, and the cosmological parameter measurements for each mock are then averaged together.}
We use the the TNS model \citep{TNS2010}, using Regularized Perturbation Theory (RegPT) at 2-loop order \citep{Taruya2012}, as described in \citet{DeMattia2020}.
An analytic Gaussian covariance matrix is used, following the method of \citet{Grieb2016}. 
\highlight{The Gaussian covariance matrix is in agreement with the EZmocks to within a level of 10\%.}
Two different sets of fits are performed. 
In the first set, the fits are applied over the $k$-range $0.03<k<0.12~\invhMpc$ for the monopole, quadrupole and hexadecapole, testing the prediction on large, linear scales. For this range we use a linear Kaiser model, with a Lorentzian Finger-of-God term.
In the second set of fits, the $k$-range is extended to $0.03<k<0.20~\invhMpc$ for all three multipoles, which tests the prediction on smaller scales, where the linear approximation
begins to break down.

We measure the growth rate, $\fsig$, and we perform the Alcock-Paczynski test \citep{Alcock1979} with the scaling parameters $\apar$ and $\aperp$. Those are related to the cosmological distances $\dH(z)/\rdrag$ and $\dM(z)/\rdrag$, at the redshift of the mocks ($z=0.876$), through
\begin{align}
\apar = \frac{\dH(z) \rdrag^\mathrm{fid}}{\dH^\mathrm{fid}(z) \rdrag}  && \mathrm{and} && \aperp = \frac{\dM(z)\rdrag^\mathrm{fid}}{\dM^\mathrm{fid}(z) \rdrag},
\end{align}
where the label `fid' indicates quantities in an assumed fiducial cosmology. 
In this analysis, we use the known cosmology of the EZmock boxes as the fiducial cosmology.

The results of the fitting procedure are shown in Table~\ref{tab:ezmock_results}.
The table shows the measured and predicted values of the 
cross covariance, $\rho^{xy}$, and the gain in the uncertainties,
$\sqrt{\sigma^\mathrm{3-los}}$, when averaging over the three lines of sight,
for the parameters $\fsig$, $\dH/\rdrag$ and $\dM/\rdrag$. 
Fisher forecasts are based on Eq.~\ref{eq:fisher_forecast_cross}. For $\vect{C}$ we take the covariance matrix used in the fits. $\vect{C}^{uv}$ is determined from Section~\ref{sec:cross_covariance_theory}, using the mean $f$, $\beta$ values obtained in the fits and $f\sigma_{d}$ as previously measured in the mocks.

The upper part of Table~\ref{tab:ezmock_results} shows the results from the first set of fits, which covers a narrow $k$-range. There is a small reduction in the uncertainties for $\dM/\rdrag$, with a slightly larger reduction for $\dH/\rdrag$, which are both in good agreement with the prediction.
The measurements for $\fsig$ show a strong anti-correlation, with errors reduced by a factor much greater than $\sqrt{1/3}$. Measurements for $\fsig$ are within $2\sigma$ of the Fisher forecasts.

When extending to smaller, quasi non-linear scales, small changes are seen in the measurements for $\dM/\rdrag$ and $\dH/\rdrag$, while a larger change is seen for $\fsig$. The measurements of $\fsig$ show a weaker anti-correlation, with a gain in the uncertainties closer to the factor of $1/\sqrt{3}$ expected for uncorrelated measurements. Fisher forecasts remain in relatively good agreement with measurements.

\highlight{
When performing these fits, it is important that the covariance matrices used are consistent with 
the measured fluctuations. We have fit the model to the power spectrum measurements for each line 
of sight individually, then taken the average of these results. This makes it straightforward 
to compute the error on the averaged values, with cross-correlations given by a Fisher forecast, 
and the covariance matrix is fully consistent with the data.
Alternatively, the model could be fit once to the average of the power spectrum measurements. 
The covariance matrix would therefore need to be changed compared to the single line-of-sight measurement as obtained from actual data.
The measured multipoles would be weighted differently by the covariance matrix, such that the model would not be tested in the same conditions as for the actual data.
}

\begin{table}
\caption{Measured values of the \highlight{cross-correlation}, $\rho^{xy}$, and the gain in uncertainties,
$\sqrt{\rho^\mathrm{3-los}}$, from the EZmocks, compared with the prediction from the Fisher 
analysis, for the parameters $\fsig$, $\dH/\rdrag$, and $\dM/\rdrag$. The upper panel is for the fits
over a small $k$-range, $0.03<k<0.12~\invhMpc$, while the lower panel shows the results for
the fits which cover a wider $k$-range, $0.03<k<0.2~\invhMpc$. The error on the measurements is estimated by splitting the 300 mocks into 6 sets of 50 mocks.}
\begin{tabular}{cccc}
\hline
$[0.03,0.12]\invhMpc$ & $\fsig$ & $\dH/\rdrag$ & $\dM/\rdrag$ \\
\hline
$\rho^{xy}$ & $-0.244 \pm 0.016$ & $0.420 \pm 0.037$ & $0.723 \pm 0.024$ \\
Prediction & -0.274 & 0.405 & 0.726 \\
$\sqrt{\rho^\mathrm{3-los}}$ & $0.414 \pm 0.013$ & $0.783 \pm 0.016$ & $0.903 \pm 0.009$\\
Prediction & 0.388 & 0.777 & 0.904 \\
\hline
\hline
$[0.03,0.2]\invhMpc$ & $\fsig$ & $\dH/\rdrag$ & $\dM/\rdrag$ \\
\hline
$\rho^{xy}$ & $-0.076 \pm 0.041$ & $0.480 \pm 0.038$ & $0.756 \pm 0.022$ \\
Prediction & -0.035 & 0.507 & 0.754 \\
$\sqrt{\rho^\mathrm{3-los}}$ & $0.532 \pm 0.025$ & $0.808 \pm 0.017$ & $0.915 \pm 0.008$\\
Prediction & 0.557 & 0.819 & 0.914 \\
\hline
\end{tabular}
\label{tab:ezmock_results}
\end{table}


\section{Conclusions}

\label{sec:conclusions}

To validate the models used in the cosmological analysis for large galaxy
surveys, and to estimate systematics, it is essential to utilize realistic 
mock catalogues. The cosmological parameters that are measured can be compared 
against the true value, since the cosmology of the mock is known.
In the eBOSS survey, mock challenges have been performed for the LRGs, ELGs
and quasars, which utilized the OuterRim N-body simulation, to assess
the modelling systematics.

As shown in \citet{Alam2020} and \citet{Smith2020}, there is scatter in the measurements 
of the cosmological parameters for different choices of the line of sight.
This is due to cosmic variance within the mock, which has a finite size. 
However, for the quasar sample, the shifts in the measurements of $\fsig$
are as large as $\sim 5\%$. For the eBOSS mock challenges, it is important that
this effect is understood, and especially important for validating models at the
high precision needed for future surveys, such as DESI.

We investigate the impact the line of sight has on the power spectrum and
correlation function measurements
from the OuterRim simulation, with observers positioned in the $x$, $y$ and 
$z$-directions. The halo catalogue used is comparable in redshift and linear bias
of the eBOSS quasar sample, where this effect is large,
but our results from this study are not specific to any individual tracer.
While the variations between the monopole measurements are small,
a much larger scatter is seen in the quadrupole. On large scales, the 
scatter between the correlation function quadrupole measurements  
is as much as $\sim 10\%$.

Since the amplitude of the quadrupole depends on the velocities, we investigate
the velocity distributions in the simulation, along each of the 3 axes. Variations in the pairwise
velocity distributions are small, resulting in differences of less than $0.1\%$
in the $DD(s,\mu)$ pair counts (for bins with $\mu$ close to 1, which are affected 
the most by velocities). These small  
variations are amplified when the correlation function is calculated. 
It is primarily the rms of the velocity distribution (in each bin of $s$)
that is responsible for the variations in the quadrupole, with a 
large rms resulting in a larger amplitude (i.e. the quadrupole is more negative).
The small variations seen in the velocity distributions are expected from
cosmic variance.

The variations in the two-point clustering measurements can be
mitigated by averaging together measurements taken using different
lines of sight. We derive an expression for the \highlight{cross-correlation} for
two power spectrum multipole measurements, assuming linear theory,
in the simplified case of orthogonal lines of sight with no shot noise
(Eqs.~\ref{eq:cross_correlation_orthogonal}-\ref{eq:auto_correlation_orthogonal_los}), 
and in the more general case of any lines of sight, including shot noise
(Eqs.~\ref{eq:cross_correlation_shot}-\ref{eq:auto_correlation_shot_los}). 
Monopole measurements are highly correlated, with a weak
correlation for the hexadecapole, while the 
quadrupole measurements are anti-correlated.

We show how the \highlight{cross-correlation} is related to the reduction in the 
variance when averaging together multiple lines of sight.
In the case of averaging together 3 orthogonal lines of sight, the reduction
is given by Eq.~\ref{eq:variance_reduction_3los}.
In the OuterRim halo catalogue, there is only a very small reduction in the variance for 
the monopole. In the limit
that there is no RSD, the variance is unchanged, since the three
measurements are identical. For the quadrupole, the anti-correlation
results in large gains in the variance, much larger than a factor of $1/3$
(or $1/\sqrt{3}$ in the uncertainties)
that would be expected if the measurements were uncorrelated.
In the limit that there is no RSD, the relative variance 
goes to 0. In the OuterRim halo catalogue, on large scales, the
variance in the combined quadrupole measurement is $\sim 25$ times smaller than the
individual measurements, resulting in uncertainties that are more than 
5 times smaller.

A reduction in the variance of the power spectrum measurements also
results in a reduction in the uncertainties on the measured cosmological
parameters. We measure this for the eBOSS ELG sample,
using a set of 300 EZmocks boxes. We measure the reduction in the
uncertainties of $\fsig$, $\dM/\rdrag$ and $\dH/\rdrag$, and compare this
with the gain predicted from performing a Fisher analysis.
When performing the fit to scales of $0.12~\invhMpc$, we expect small gains 
in the uncertainties on $\dM/\rdrag$ and $\dH/\rdrag$ of $\sim 0.90$ and $\sim 0.77$, 
respectively, with much larger gain in $\fsig$ of $\sim 0.39$ due to the measurements being
strongly anti-correlated. 
Extending the fit to smaller scales of $0.2~\invhMpc$, the anti-correlation for $\fsig$ is reduced,
resulting in gains that are only slightly better than a factor of $1/\sqrt{3}$.
The Fisher forecasts show reasonable agreement with the measurements from the mocks.

\highlight{In this work, we have only considered cubic boxes, where we have used the plane parallel 
approximation. For mocks that cover a wide area, where the lines of sight to each object are no
longer parallel, it will still be possible to place observers in different positions, and average the
clustering or cosmological measurements to reduce the uncertainties.}

By averaging together the clustering measurements from the 3 orthogonal lines of sight, 
large gains can be made in the variance of the quadrupole, and hence
$\fsig$ measurements, which are better than a factor of $1/3$. In mock
challenges, when validating models, it is therefore important to average these
measurements together. This is particularly true if a single mock is used, 
which was the case for the OuterRim mocks used in eBOSS. 
The effect is stronger for tracers, such as quasars, which are more strongly biased,
and have a smaller value of $\beta$.
For a set of many mock catalogues, where the total volume is large,
it is still worthwhile to combine the results measured
from multiple lines of sight. This will allow tighter constraints to be placed on the models,
without needing to generate more mocks. 

\section*{Acknowledgements}

AS acknowledges support from grant ANR-16-CE31-0021, eBOSS and 
ANR-17-CE31-0024-01, NILAC.

Funding for the Sloan Digital Sky Survey IV has been provided by the Alfred P. Sloan Foundation, the U.S. Department of Energy Office of Science, and the Participating Institutions. SDSS-IV acknowledges
support and resources from the Center for High-Performance Computing at
the University of Utah. The SDSS web site is www.sdss.org.
In addition, this research relied on resources provided to the eBOSS
Collaboration by the National Energy Research Scientific Computing
Center (NERSC).  NERSC is a U.S. Department of Energy Office of Science
User Facility operated under Contract No. DE-AC02-05CH11231.

SDSS-IV is managed by the Astrophysical Research Consortium for the 
Participating Institutions of the SDSS Collaboration including the 
Brazilian Participation Group, the Carnegie Institution for Science, 
Carnegie Mellon University, the Chilean Participation Group, the French Participation Group, Harvard-Smithsonian Center for Astrophysics, 
Instituto de Astrof\'isica de Canarias, The Johns Hopkins University, Kavli Institute for the Physics and Mathematics of the Universe (IPMU) / 
University of Tokyo, the Korean Participation Group, Lawrence Berkeley National Laboratory, 
Leibniz Institut f\"ur Astrophysik Potsdam (AIP),  
Max-Planck-Institut f\"ur Astronomie (MPIA Heidelberg), 
Max-Planck-Institut f\"ur Astrophysik (MPA Garching), 
Max-Planck-Institut f\"ur Extraterrestrische Physik (MPE), 
National Astronomical Observatories of China, New Mexico State University, 
New York University, University of Notre Dame, 
Observat\'ario Nacional / MCTI, The Ohio State University, 
Pennsylvania State University, Shanghai Astronomical Observatory, 
United Kingdom Participation Group,
Universidad Nacional Aut\'onoma de M\'exico, University of Arizona, 
University of Colorado Boulder, University of Oxford, University of Portsmouth, 
University of Utah, University of Virginia, University of Washington, University of Wisconsin, 
Vanderbilt University, and Yale University.

This work used the DiRAC@Durham facility managed by the Institute for Computational Cosmology on behalf of the STFC DiRAC HPC Facility (www.dirac.ac.uk). The equipment was funded by BEIS capital funding via STFC capital grants ST/K00042X/1, ST/P002293/1, ST/R002371/1 and ST/S002502/1, Durham University and STFC operations grant ST/R000832/1. DiRAC is part of the National e-Infrastructure.

\section*{Data availability}

The OuterRim simulation snapshot used in this work is publicly available at \url{https://cosmology.alcf.anl.gov/outerrim}.
The EZmock catalogues and their clustering measurements will be shared on reasonable request to the corresponding author with permission of Cheng Zhao.




\bibliographystyle{mnras}
\bibliography{ref} 

\begin{thebibliography}{}
\makeatletter
\relax
\def\mn@urlcharsother{\let\do\@makeother \do\$\do\&\do\#\do\^\do\_\do\%\do\~}
\def\mn@doi{\begingroup\mn@urlcharsother \@ifnextchar [ {\mn@doi@}
  {\mn@doi@[]}}
\def\mn@doi@[#1]#2{\def\@tempa{#1}\ifx\@tempa\@empty \href
  {http://dx.doi.org/#2} {doi:#2}\else \href {http://dx.doi.org/#2} {#1}\fi
  \endgroup}
\def\mn@eprint#1#2{\mn@eprint@#1:#2::\@nil}
\def\mn@eprint@arXiv#1{\href {http://arxiv.org/abs/#1} {{\tt arXiv:#1}}}
\def\mn@eprint@dblp#1{\href {http://dblp.uni-trier.de/rec/bibtex/#1.xml}
  {dblp:#1}}
\def\mn@eprint@#1:#2:#3:#4\@nil{\def\@tempa {#1}\def\@tempb {#2}\def\@tempc
  {#3}\ifx \@tempc \@empty \let \@tempc \@tempb \let \@tempb \@tempa \fi \ifx
  \@tempb \@empty \def\@tempb {arXiv}\fi \@ifundefined
  {mn@eprint@\@tempb}{\@tempb:\@tempc}{\expandafter \expandafter \csname
  mn@eprint@\@tempb\endcsname \expandafter{\@tempc}}}

\bibitem[\protect\citeauthoryear{{Alam} et~al.,}{{Alam}
  et~al.}{2020}]{Alam2020}
{Alam} S.,  et~al., 2020, arXiv e-prints, \href
  {https://ui.adsabs.harvard.edu/abs/2020arXiv200709004A} {p. arXiv:2007.09004}

\bibitem[\protect\citeauthoryear{{Alcock} \& {Paczynski}}{{Alcock} \&
  {Paczynski}}{1979}]{Alcock1979}
{Alcock} C.,  {Paczynski} B.,  1979, \mn@doi [\nat] {10.1038/281358a0}, \href
  {https://ui.adsabs.harvard.edu/abs/1979Natur.281..358A} {281, 358}

\bibitem[\protect\citeauthoryear{{Bianchi}, {Chiesa}  \& {Guzzo}}{{Bianchi}
  et~al.}{2015}]{Bianchi2015}
{Bianchi} D.,  {Chiesa} M.,   {Guzzo} L.,  2015, \mn@doi [\mnras]
  {10.1093/mnras/stu2080}, \href
  {https://ui.adsabs.harvard.edu/abs/2015MNRAS.446...75B} {446, 75}

\bibitem[\protect\citeauthoryear{{Blanton} et~al.,}{{Blanton}
  et~al.}{2017}]{Blanton2017}
{Blanton} M.~R.,  et~al., 2017, \mn@doi [\aj] {10.3847/1538-3881/aa7567}, \href
  {https://ui.adsabs.harvard.edu/abs/2017AJ....154...28B} {154, 28}

\bibitem[\protect\citeauthoryear{{Chuang}, {Kitaura}, {Prada}, {Zhao}  \&
  {Yepes}}{{Chuang} et~al.}{2015}]{Chuang2015}
{Chuang} C.-H.,  {Kitaura} F.-S.,  {Prada} F.,  {Zhao} C.,   {Yepes} G.,  2015,
  \mn@doi [\mnras] {10.1093/mnras/stu2301}, \href
  {http://adsabs.harvard.edu/abs/2015MNRAS.446.2621C} {446, 2621}

\bibitem[\protect\citeauthoryear{{Cole} et~al.,}{{Cole}
  et~al.}{2005}]{Cole2005}
{Cole} S.,  et~al., 2005, \mn@doi [\mnras] {10.1111/j.1365-2966.2005.09318.x},
  \href {https://ui.adsabs.harvard.edu/abs/2005MNRAS.362..505C} {362, 505}

\bibitem[\protect\citeauthoryear{{Cuesta-Lazaro}, {Li}, {Eggemeier}, {Zarrouk},
  {Baugh}, {Nishimichi}  \& {Takada}}{{Cuesta-Lazaro}
  et~al.}{2020}]{Cuesta2020}
{Cuesta-Lazaro} C.,  {Li} B.,  {Eggemeier} A.,  {Zarrouk} P.,  {Baugh} C.~M.,
  {Nishimichi} T.,   {Takada} M.,  2020, arXiv e-prints, \href
  {https://ui.adsabs.harvard.edu/abs/2020arXiv200202683C} {p. arXiv:2002.02683}

\bibitem[\protect\citeauthoryear{{DESI Collaboration} et~al.,}{{DESI
  Collaboration} et~al.}{2016a}]{DESI2016a}
{DESI Collaboration} et~al., 2016a, arXiv e-prints, \href
  {https://ui.adsabs.harvard.edu/abs/2016arXiv161100036D} {p. arXiv:1611.00036}

\bibitem[\protect\citeauthoryear{{DESI Collaboration} et~al.,}{{DESI
  Collaboration} et~al.}{2016b}]{DESI2016b}
{DESI Collaboration} et~al., 2016b, arXiv e-prints, \href
  {https://ui.adsabs.harvard.edu/abs/2016arXiv161100037D} {p. arXiv:1611.00037}

\bibitem[\protect\citeauthoryear{{Davis}, {Efstathiou}, {Frenk}  \&
  {White}}{{Davis} et~al.}{1985}]{Davis1985}
{Davis} M.,  {Efstathiou} G.,  {Frenk} C.~S.,   {White} S.~D.~M.,  1985,
  \mn@doi [\apj] {10.1086/163168}, \href
  {http://adsabs.harvard.edu/abs/1985ApJ...292..371D} {292, 371}

\bibitem[\protect\citeauthoryear{{Dawson} et~al.,}{{Dawson}
  et~al.}{2016}]{Dawson2016}
{Dawson} K.~S.,  et~al., 2016, \mn@doi [\aj] {10.3847/0004-6256/151/2/44},
  \href {https://ui.adsabs.harvard.edu/abs/2016AJ....151...44D} {151, 44}

\bibitem[\protect\citeauthoryear{{Eisenstein} et~al.,}{{Eisenstein}
  et~al.}{2005}]{Eisenstein2005}
{Eisenstein} D.~J.,  et~al., 2005, \mn@doi [\apj] {10.1086/466512}, \href
  {https://ui.adsabs.harvard.edu/abs/2005ApJ...633..560E} {633, 560}

\bibitem[\protect\citeauthoryear{{Fisher}}{{Fisher}}{1995}]{Fisher1995}
{Fisher} K.~B.,  1995, \mn@doi [\apj] {10.1086/175980}, \href
  {https://ui.adsabs.harvard.edu/abs/1995ApJ...448..494F} {448, 494}

\bibitem[\protect\citeauthoryear{{Grieb}, {S{\'a}nchez}, {Salazar-Albornoz}  \&
  {Dalla Vecchia}}{{Grieb} et~al.}{2016}]{Grieb2016}
{Grieb} J.~N.,  {S{\'a}nchez} A.~G.,  {Salazar-Albornoz} S.,   {Dalla Vecchia}
  C.,  2016, \mn@doi [\mnras] {10.1093/mnras/stw065}, \href
  {https://ui.adsabs.harvard.edu/abs/2016MNRAS.457.1577G} {457, 1577}

\bibitem[\protect\citeauthoryear{{Guzzo} et~al.,}{{Guzzo}
  et~al.}{2008}]{Guzzo2008}
{Guzzo} L.,  et~al., 2008, \mn@doi [\nat] {10.1038/nature06555}, \href
  {https://ui.adsabs.harvard.edu/abs/2008Natur.451..541G} {451, 541}

\bibitem[\protect\citeauthoryear{{Habib} et~al.,}{{Habib}
  et~al.}{2016}]{Habib2016}
{Habib} S.,  et~al., 2016, \mn@doi [\na] {10.1016/j.newast.2015.06.003}, \href
  {http://adsabs.harvard.edu/abs/2016NewA...42...49H} {42, 49}

\bibitem[\protect\citeauthoryear{{Hand}, {Feng}, {Beutler}, {Li}, {Modi},
  {Seljak}  \& {Slepian}}{{Hand} et~al.}{2018}]{Hand2018}
{Hand} N.,  {Feng} Y.,  {Beutler} F.,  {Li} Y.,  {Modi} C.,  {Seljak} U.,
  {Slepian} Z.,  2018, \mn@doi [\aj] {10.3847/1538-3881/aadae0}, \href
  {https://ui.adsabs.harvard.edu/abs/2018AJ....156..160H} {156, 160}

\bibitem[\protect\citeauthoryear{{Heitmann} et~al.,}{{Heitmann}
  et~al.}{2019a}]{Heitmann2019a}
{Heitmann} K.,  et~al., 2019a, arXiv e-prints, \href
  {http://adsabs.harvard.edu/abs/2019arXiv190411970H} {}

\bibitem[\protect\citeauthoryear{{Heitmann} et~al.,}{{Heitmann}
  et~al.}{2019b}]{Heitmann2019b}
{Heitmann} K.,  et~al., 2019b, \mn@doi [\apjs] {10.3847/1538-4365/ab3724},
  \href {https://ui.adsabs.harvard.edu/abs/2019ApJS..244...17H} {244, 17}

\bibitem[\protect\citeauthoryear{{Howlett} \& {Percival}}{{Howlett} \&
  {Percival}}{2017}]{Howlett2017}
{Howlett} C.,  {Percival} W.~J.,  2017, \mn@doi [\mnras]
  {10.1093/mnras/stx2342}, \href
  {https://ui.adsabs.harvard.edu/abs/2017MNRAS.472.4935H} {472, 4935}

\bibitem[\protect\citeauthoryear{{Jackson}}{{Jackson}}{1972}]{Jackson1972}
{Jackson} J.~C.,  1972, \mn@doi [\mnras] {10.1093/mnras/156.1.1P}, \href
  {https://ui.adsabs.harvard.edu/abs/1972MNRAS.156P...1J} {156, 1P}

\bibitem[\protect\citeauthoryear{{Juszkiewicz}, {Fisher}  \&
  {Szapudi}}{{Juszkiewicz} et~al.}{1998}]{Juszkiewicz1998}
{Juszkiewicz} R.,  {Fisher} K.~B.,   {Szapudi} I.,  1998, \mn@doi [\apjl]
  {10.1086/311558}, \href
  {https://ui.adsabs.harvard.edu/abs/1998ApJ...504L...1J} {504, L1}

\bibitem[\protect\citeauthoryear{{Kaiser}}{{Kaiser}}{1987}]{Kaiser1987}
{Kaiser} N.,  1987, \mn@doi [\mnras] {10.1093/mnras/227.1.1}, \href
  {https://ui.adsabs.harvard.edu/abs/1987MNRAS.227....1K} {227, 1}

\bibitem[\protect\citeauthoryear{{Komatsu} et~al.,}{{Komatsu}
  et~al.}{2011}]{Komatsu2011}
{Komatsu} E.,  et~al., 2011, \mn@doi [\apjs] {10.1088/0067-0049/192/2/18},
  \href {http://adsabs.harvard.edu/abs/2011ApJS..192...18K} {192, 18}

\bibitem[\protect\citeauthoryear{{Kuruvilla} \& {Porciani}}{{Kuruvilla} \&
  {Porciani}}{2018}]{Kuruvilla2018}
{Kuruvilla} J.,  {Porciani} C.,  2018, \mn@doi [\mnras]
  {10.1093/mnras/sty1654}, \href
  {https://ui.adsabs.harvard.edu/abs/2018MNRAS.479.2256K} {479, 2256}

\bibitem[\protect\citeauthoryear{{Landy} \& {Szalay}}{{Landy} \&
  {Szalay}}{1993}]{Landy1993}
{Landy} S.~D.,  {Szalay} A.~S.,  1993, \mn@doi [\apj] {10.1086/172900}, \href
  {https://ui.adsabs.harvard.edu/abs/1993ApJ...412...64L} {412, 64}

\bibitem[\protect\citeauthoryear{{Linder}}{{Linder}}{2005}]{Linder2005}
{Linder} E.~V.,  2005, \mn@doi [\prd] {10.1103/PhysRevD.72.043529}, \href
  {https://ui.adsabs.harvard.edu/abs/2005PhRvD..72d3529L} {72, 043529}

\bibitem[\protect\citeauthoryear{{Meiksin} \& {White}}{{Meiksin} \&
  {White}}{1999}]{Meiksin1999}
{Meiksin} A.,  {White} M.,  1999, \mn@doi [\mnras]
  {10.1046/j.1365-8711.1999.02825.x}, \href
  {https://ui.adsabs.harvard.edu/abs/1999MNRAS.308.1179M} {308, 1179}

\bibitem[\protect\citeauthoryear{{Percival} \& {White}}{{Percival} \&
  {White}}{2009}]{Percival2009}
{Percival} W.~J.,  {White} M.,  2009, \mn@doi [\mnras]
  {10.1111/j.1365-2966.2008.14211.x}, \href
  {https://ui.adsabs.harvard.edu/abs/2009MNRAS.393..297P} {393, 297}

\bibitem[\protect\citeauthoryear{{Planck Collaboration} et~al.,}{{Planck
  Collaboration} et~al.}{2014}]{Planck2014}
{Planck Collaboration} et~al., 2014, \mn@doi [\aap]
  {10.1051/0004-6361/201321591}, \href
  {https://ui.adsabs.harvard.edu/abs/2014A&A...571A..16P} {571, A16}

\bibitem[\protect\citeauthoryear{{Pueblas} \& {Scoccimarro}}{{Pueblas} \&
  {Scoccimarro}}{2009}]{Pueblas2009}
{Pueblas} S.,  {Scoccimarro} R.,  2009, \mn@doi [\prd]
  {10.1103/PhysRevD.80.043504}, \href
  {https://ui.adsabs.harvard.edu/abs/2009PhRvD..80d3504P} {80, 043504}

\bibitem[\protect\citeauthoryear{{Reid} \& {White}}{{Reid} \&
  {White}}{2011}]{Reid2011}
{Reid} B.~A.,  {White} M.,  2011, \mn@doi [\mnras]
  {10.1111/j.1365-2966.2011.19379.x}, \href
  {https://ui.adsabs.harvard.edu/abs/2011MNRAS.417.1913R} {417, 1913}

\bibitem[\protect\citeauthoryear{{Rossi} et~al.,}{{Rossi}
  et~al.}{2020}]{Rossi2020}
{Rossi} G.,  et~al., 2020, arXiv e-prints, \href
  {https://ui.adsabs.harvard.edu/abs/2020arXiv200709002R} {p. arXiv:2007.09002}

\bibitem[\protect\citeauthoryear{{Scoccimarro}}{{Scoccimarro}}{2004}]{Scoccimarro2004}
{Scoccimarro} R.,  2004, \mn@doi [\prd] {10.1103/PhysRevD.70.083007}, \href
  {https://ui.adsabs.harvard.edu/abs/2004PhRvD..70h3007S} {70, 083007}

\bibitem[\protect\citeauthoryear{{Smith} et~al.,}{{Smith}
  et~al.}{2020}]{Smith2020}
{Smith} A.,  et~al., 2020, arXiv e-prints, \href
  {https://ui.adsabs.harvard.edu/abs/2020arXiv200709003S} {p. arXiv:2007.09003}

\bibitem[\protect\citeauthoryear{{Taruya}, {Nishimichi}  \& {Saito}}{{Taruya}
  et~al.}{2010a}]{Taruya2010:1006.0699v1}
{Taruya} A.,  {Nishimichi} T.,   {Saito} S.,  2010a, \mn@doi [\prd]
  {10.1103/PhysRevD.82.063522}, \href
  {https://ui.adsabs.harvard.edu/abs/2010PhRvD..82f3522T} {82, 063522}

\bibitem[\protect\citeauthoryear{{Taruya}, {Nishimichi}  \& {Saito}}{{Taruya}
  et~al.}{2010b}]{TNS2010}
{Taruya} A.,  {Nishimichi} T.,   {Saito} S.,  2010b, \mn@doi [\prd]
  {10.1103/PhysRevD.82.063522}, \href
  {http://adsabs.harvard.edu/abs/2010PhRvD..82f3522T} {82, 063522}

\bibitem[\protect\citeauthoryear{{Taruya}, {Bernardeau}, {Nishimichi}  \&
  {Codis}}{{Taruya} et~al.}{2012}]{Taruya2012}
{Taruya} A.,  {Bernardeau} F.,  {Nishimichi} T.,   {Codis} S.,  2012, \mn@doi
  [\prd] {10.1103/PhysRevD.86.103528}, \href
  {https://ui.adsabs.harvard.edu/abs/2012PhRvD..86j3528T} {86, 103528}

\bibitem[\protect\citeauthoryear{{Tinker}}{{Tinker}}{2007}]{Tinker2007}
{Tinker} J.~L.,  2007, \mn@doi [\mnras] {10.1111/j.1365-2966.2006.11157.x},
  \href {https://ui.adsabs.harvard.edu/abs/2007MNRAS.374..477T} {374, 477}

\bibitem[\protect\citeauthoryear{{Wang}, {Reid}  \& {White}}{{Wang}
  et~al.}{2014}]{Wang2014}
{Wang} L.,  {Reid} B.,   {White} M.,  2014, \mn@doi [\mnras]
  {10.1093/mnras/stt1916}, \href
  {https://ui.adsabs.harvard.edu/abs/2014MNRAS.437..588W} {437, 588}

\bibitem[\protect\citeauthoryear{{Zhao} et~al.,}{{Zhao}
  et~al.}{2020}]{Zhao2020}
{Zhao} C.,  et~al., 2020, arXiv e-prints, \href
  {https://ui.adsabs.harvard.edu/abs/2020arXiv200708997Z} {p. arXiv:2007.08997}

\bibitem[\protect\citeauthoryear{{de Mattia} et~al.,}{{de Mattia}
  et~al.}{2020}]{DeMattia2020}
{de Mattia} A.,  et~al., 2020, arXiv e-prints, \href
  {https://ui.adsabs.harvard.edu/abs/2020arXiv200709008D} {p. arXiv:2007.09008}

\bibitem[\protect\citeauthoryear{{eBOSS Collaboration} et~al.,}{{eBOSS
  Collaboration} et~al.}{2020}]{eBOSS_Cosmology2020}
{eBOSS Collaboration} et~al., 2020, arXiv e-prints, \href
  {https://ui.adsabs.harvard.edu/abs/2020arXiv200708991E} {p. arXiv:2007.08991}

\makeatother
\end{thebibliography}








\bsp	
\label{lastpage}
\end{document}